\documentclass{article}
\usepackage{graphicx} % Required for inserting images

\usepackage[a4paper,top=2cm,bottom=2cm,left=3cm,right=3cm,marginparwidth=1.75cm]{geometry}

\usepackage{ulem} %\sout command

\usepackage{amsmath}
\usepackage{amssymb}
\usepackage{tikz-cd} 
\usetikzlibrary{fadings}
\usepackage{fancyvrb} % Needed for verbatim in footnotes
\usepackage{amsthm}
\usepackage{array}

\usepackage[colorlinks=true, allcolors=blue]{hyperref}
\usepackage{bm}
\usepackage{makecell}
\usepackage{subcaption}
\usepackage{tikz}

\usepackage{cite}%citationrange

\VerbatimFootnotes  % Needed for verbatim in footnotes

\newcommand{\nflux}{N_{\textrm{flux}}}

\theoremstyle{remark}

\theoremstyle{definition}

\renewcommand{\Re}{\textrm{Re}}
\renewcommand{\Im}{\textrm{Im}}

\newcommand{\Xx}{$\{X,{\bf{x}}\}$}

\definecolor{airforceblue}{rgb}{0.36, 0.54, 0.66}

\vspace{1cm}

\title{       {\Large \bf Sampling String Vacua Using Generative Models}}

\vspace{2cm}

\author{
Moritz Walden\footnote{Corresponding author; moritz.walden@physics.uu.se},
 Magdalena Larfors\footnote{magdalena.larfors@physics.uu.se} }

\date{\today}

\begin{document}

\maketitle
\begin{center} {{\it Department of Physics and Astronomy \& Department of Mathematics, 
\\Uppsala University,\\
       L\"agerhyddsv. 1, SE-751 20 Uppsala, Sweden.}}\\

\end{center}

\begin{abstract}
\noindent
We apply generative models to a key problem in the string compactification program, namely construction of type IIB string vacua. To this end, we make use of a Bayesian Flow Network, a generative model capable of handling discrete data, to generate flux vectors that give rise to type IIB vacua. Furthermore, we sample flux vacua that have certain desirable properties by employing a Transformer as a conditional generative model. Both models demonstrate good performance in finding flux vacua and thus prove to be powerful tools in the exploration of the string landscape.

\end{abstract}

\thispagestyle{empty}
\setcounter{page}{0}
\newpage

\tableofcontents
\clearpage

\section{Introduction}
\label{sec:intro}

String theory is a leading proposal for quantum gravity.  Compactifying string theory on some miniscule, compact space allows us to connect the theory with four-dimensional physics. This procedure famously leads to a vast landscape of string theory vacua.\footnote{Their magnitude has been estimated in the type IIB and M-theory setting in \cite{Ashok:2003gk,Acharya:2006zw}, F-theory vacua have been counted in \cite{Taylor:2015xtz}, type IIA vacua in \cite{Loges:2022mao} and vacua obtained from heterotic string theory have been enumerated in \cite{Constantin:2018xkj}.} A key problem in string phenomenology is that most of these vacua are expected to eventually fail to meet observational constraints, such as a particle spectrum matching that of the standard model, or a de Sitter cosmology with small positive cosmological constant.

The vastness of the string theory landscape results from the large number of suitable compact  manifolds $X$, and the freedom in choosing certain integer quantities $\bf{x} \in \mathbb{Z}^n$ that describe quantized objects that may fill the compact space. The complexity of string theory, and the intricate geometry of compact, high-dimensional, manifolds, renders the problem of identifying tuples \Xx ~that yield realistic vacua  computationally complex; leading compactification scenarios include both NP-complete \cite{Denef:2006ad,Cvetic:2010ky} and NP-hard problems \cite{Halverson:2018cio}. Similarly, the enormity of the  landscape means that systematic explorations are only possible in highly restrictive settings.

On the other hand, one may still proceed by rejection sampling: select $X$ and $\bf{x}$, and then check if a realistic string vacuum arises. If all relevant constraints are satisfied, \Xx ~can be saved for further study; otherwise the sample is rejected and one starts over. Since realistic vacua are rare in the string landscape, this method will be highly inefficient, unless we have a clever way of selecting \Xx. One approach to find such selection strategies is using machine learning (ML).\footnote{This idea is not new; in the past decade the string theory landscape has successfully been explored with a variety of supervised and unsupervised ML tools, in setting such as F-theory \cite{Bies:2020gvf}, heterotic line bundle \cite{Deen:2020dlf} and type IIB \cite{Arias-Tamargo:2022qgb} compactifications, as well as related geometrical questions \cite{Carifio:2017bov}. Further applications of Reinforcement Learning include studies in the setting of type IIA \cite{Halverson:2019tkf}, type IIB \cite{Krippendorf:2022gcl, Carta:2025asr} and heterotic line bundle \cite{Larfors:2020ugo} compactifications. Genetic Algorithms have been successfully utilized to study heterotic line bundle \cite{Abel:2021rrj, Abel:2023zwg} and type IIA  \cite{Loges:2021hvn} compactifications, as well as related geometrical questions \cite{Ruehle:2017mzq}. We refer the reader to the reviews 
\cite{He:2023csq,Ruehle:2020jrk,Anderson:2023viv,Halverson:2024hax} for a more thorough introduction to the field.}

The success of ML models relies on the hypothesis that the string landscape has structure, which can be captured by a non-trivial probability distribution of pairs \Xx ~that define realistic string vacua. Keeping $X$ fixed, we denote this  distribution by
\begin{equation} \label{eq:pvacua}
    p_{\rm vacua}(\mathbf{x}) \,.
\end{equation}
Constructing a realistic string vacuum on $X$ can then be seen as sampling from this probability distribution. Neural networks are very good at learning (approximations of) such distributions, and use this information for classification or regression tasks.  Alternatively, in generative deep learning models, the goal is to generate new data samples. More precisely, a generative model, parametrized by $\bm{\theta}$, aims to learn an approximation 
    \begin{equation}
        p_{\bm{\theta}}(\mathbf{x})
    \end{equation}
    for the (unknown) data distribution. Training (or optimization) of the model, from observation of a large number of data points, proceeds by tuning the model parameters $\bm{\theta}$ so that 
    \begin{equation}
        p_{\bm{\theta}}(\mathbf{x}) \approx p_{\rm data}(\mathbf{x}) \,.
    \end{equation}
After training, the model can be used to sample $\mathbf{x}$ from the optimal $p_{\bm{\theta}}(\mathbf{x})$, thus generating new data. Since the method is probabilistic, it is not guaranteed that the generated sample is accurate. A measure of sample accuracy is needed, and what this is will depend on the type of data.

Adapting this methodology to string phenomenology is then straight-forward. We start with a data set of valid\footnote{We will define ``valid'' in Sec.~\ref{sec:bgphys}; these vacua satisfy weaker constraints than those of realistic string vacua.} string vacua on $X$, which we view as samples from  some unknown distribution $p_{\rm vacua}(\mathbf{x})$. We select a generative model, and train it, using the data set, to learn an approximation of  $p_{\rm vacua}(\mathbf{x})$. We check the sample accuracy of the model by checking that the generated \Xx ~obey all constraints for valid vacua; this exact check can be done numerically, to some desired precision. Once the vacua are generated and checked, they can be used for any downstream task in string phenomenology. Or, the trained models can be used in conjunction with other computational tools, that rely on sampling $\mathbf{x}$.

In this paper, we will explore flux vacua of type IIB string theory, for which $X$ is a Calabi-Yau orientifold. This is a popular setting for cosmology models in string theory \cite{Kachru:2003aw,Balasubramanian:2005zx},\footnote{See \cite{McAllister:2024lnt} for the state of the art of these constructions.} and the vacua are known to have interesting structural properties.\footnote{The non-trivial structure of the type IIB landscape is well-documented, e.g.~via approximate statistical distributions \cite{Ashok:2003gk, Denef:2004ze} and direct numerical \cite{Danielsson:2006xw,Chialva:2007sv} approaches. The statistical structure of M-theory is studied in \cite{Douglas:2003um} and \cite{Krippendorf:2022gcl} explores the structure of a simplified landscape model.} Early this year, Chauhan et.al.~\cite{Chauhan:2025rdj} developed an algorithm (leveraging \cite{Demirtas:2022hqf,Dubey:2023dvu}) for the construction of such vacua in a designated, finite region of moduli space. Applying this algorithm to the CY manifold $\mathbb{CP}^4_{[11169]}[18]$, Chauhan et.al.~systematically constructed flux vacua in the given region of moduli space. 
In this paper, we will use one of the datasets of \cite{Chauhan:2025rdj} to train generative models to predict a vector $\bf{x}$ of flux integers that leads to valid string vacua in this setting.  

Our study requires the selection of a generative model. There is, by now, a variety of such models, including Generative Adversarial Networks \cite{2014arXiv1406.2661G}, Variational Autoencoders (VAE) \cite{2013arXiv1312.6114K,2014arXiv1401.4082J}, Diffusion Models \cite{2015arXiv150303585S} and Transformers \cite{vaswani2023attentionneed}. 
While some generative architectures have already been tested on string-related problems, such as GANs \cite{Erbin:2018csv, Halverson:2020opj}, Transformers \cite{halverson2025learningtopologicalinvariance, Yip:2025hon} and CVAEs \cite{Krippendorf:2025mhp, Seong:2024wkt}, the topic is far from being exhausted. 
In this paper, we employ two types of generative models, Bayesian Flow Networks (BFNs) \cite{2023arXiv230807037G} and Transformers \cite{vaswani2023attentionneed}. To the best of our knowledge, this is the first time that BFNs have been applied to string theory, or any mathematical problem. The methodological similarities and differences between BFNs and Transformers motivate a study that compares their performance and accuracy for string theory problems.

 This is a method paper where we test the abilities of BFNs and Transformers to generate new\footnote{New here means that the vacuum is not present in the training set.} type IIB flux vacua in a pre-designated regime of moduli space. First, we compare the distributions of sampled vacua, both over the field space and with respect to physical properties,  with the observations made in \cite{Chauhan:2025rdj}. We also test the interpolation and extrapolation ability of the methods, i.e. how efficiently they can generate new vacua that lie in and outside of training domain.  Second, we explore sampling abilities conditioned on pre-determined physics-motivated constraints (the value of $\nflux$ and $|W_0|$).

The outline of this paper is as follows. We start, in Sec. \ref{sec:bgphys}, with a short review of type IIB flux vacua. In Sec. \ref{sec:bgml}, we describe the key properties of the generative models that will be used for sampling of flux vacua. Sec. \ref{sec:expIIB} presents the experiments performed. Here, we also display that the generative models succeed in reproducing the data distribution, extrapolate beyond the training data (BFN), and generate vacua with pre-selected properties (Transformer). We conclude with a discussion of our results and outlook to future work in Sec. \ref{sec:d&o}. Two appendices provide further explanations and technical details about the methods used, and  additional experimental results.

During the final stage of this project, Krippendorf and Liu reported on similar experiments with generative models for type IIB vacua \cite{Krippendorf:2025mhp}, which has an overlap with our study. The primary difference between the two papers lie in the choice of generative methods. We train BFNs and Transformers, with their respective standard loss functions, to sample flux vectors, as will be described in detail below.  In contrast, \cite{Krippendorf:2025mhp} employed Conditional VAE, with a custom loss function, for conditional sampling of flux vectors. The difference in architecture and training between the three models is significant. On the other hand, all three methods are generative and probabilistic, and require a post-sampling check of the validity of the sampled flux vectors.  A secondary, minor, difference between the two papers lies in the geometric settings explored. Despite the differences, our results agree with the findings of \cite{Krippendorf:2025mhp}; generative models provide a highly promising approach for the generation of string vacua, and together the papers illustrate the strengths  of the respective methods.

\section{Type IIB flux vacua}
\label{sec:bgphys}

In this paper, we perform discrete sampling experiments on flux vectors ${\bf{x}}$ for ISD flux vacua of type IIB string theory, using datasets generated for the reference \cite{Chauhan:2025rdj}. In this section, which can safely be skipped by the expert reader, we briefly recapitulate how such type IIB flux vacua  are constructed, and emphasize a few key points on the physical theory and underlying geometry. This serves to fix our conventions (which follow \cite{Chauhan:2025rdj}%with minor modifications \ml{if so}
), and specify the equations that need to be checked in order compute the sample accuracy of our experiments. More extensive reviews of flux compactifications can be found in \cite{Grana:2005jc,Douglas:2006es}. A short, mathematical review of the string compactification setting is given in \cite{Douglas:2023yof}.

\subsection{Type IIB supergravity}

 Our starting point is the low-energy supergravity limit of type IIB string theory.\footnote{See the textbook \cite{Polchinski:1998rr} for a review.} This is a 10D chiral, $SL(2,\mathbb{Z})$-invariant, theory, with fermionic and bosonic fields.  Among the latter, we will pay particular attention to the axio-dilaton, 
\[
\tau =c_0 +i s\, ,
\]
and the 3-form field strengths $F_3, H_3$.  

Compactification of type IIB string theory on a Calabi-Yau manifold $X$ with O3/O7 orientifolds leads to an effective ${\cal N}=1$ supergravity in four dimensions. This 4d theory admits many different ground states, or vacua, with negative and zero cosmological constant.\footnote{The construction of vacua with positive cosmological constant is more elaborate and will not be a focus here.} The multitude of vacua is associated to different choices of quantized fluxes, i.e. vacuum expectation values, along internal 3-cycles $C_i \in H_3(X,\mathbb{Z})$, for the 3-form field strengths:  
\[
f_i = \int_{C_i} F_3 \, , \;  h_i = \int_{C_i} H_3 \,.
\]
These fluxes lead to an effective potential that depends on the 3-cycle volume, and a vacuum requires that this potential is minimized. We collect the quantized fluxes in a flux vector 
\[
{\bf{x}}=(f,h) \;.
\]
Clearly, one may choose the flux vectors in infinitely many ways, but, as we will see momentarily, there are constraints that bound the number of flux vacua to a finite, but very large, value.

More precisely, the 4D ${\cal N}=1$ supergravity theory is governed by a real Kähler potential $K$ and a holomorphic superpotential $W$. Focusing on the complex structure moduli $z^i$ and axio-dilaton $\tau$, $K$ is given by
\begin{equation}
    K = - \ln \left(-i \Pi^\dagger \cdot \Sigma  \cdot \Pi \right)- \ln \left( -i(\tau-\bar{\tau})\right) \,,
\end{equation}
where $\Sigma$ is a symplectic intersection matrix 
and $\Pi_i(z)$ are the complex-structure dependent periods of the CY $X$. These geometric objects will be described in more detail below.

The superpotential for the theory is the  Gukov-Vafa-Witten (GVW) superpotential \cite{Gukov:1999ya}
\begin{equation} \label{eq:superpot}
    W = \int_X (F_3 - \tau H_3) \wedge \Omega = (f-\tau h)^T \cdot \Sigma \cdot \Pi(z) \, .
\end{equation}
Together, $K$ and $W$ define a scalar potential. Requiring the F-terms to vanish, namely
\begin{equation}
    \begin{split} \label{eq:Fterms}
          &D_\tau W = \frac{1}{\tau-\bar{\tau}}(f-\bar{\tau}h)^T \cdot \Sigma \cdot \Pi(z) = 0\,,  \\
          &D_i W = (f-\tau h)^T \cdot \Sigma \cdot (\partial_i + \partial_i K)\Pi(z) = 0 \,,
    \end{split}
\end{equation}
leads to minima of this scalar potential, and a candidate for a supersymmetric vacuum solution.\footnote{We neglect the K\"ahler moduli, which must also be stabilised in type IIB vacua.} Since these F-term conditions correspond to an imaginary self-dual constraint for the flux vector ${\bf{x}}=(f,h)$, such minima are referred to as ISD vacua \cite{Giddings:2001yu}. 

In addition to the F-term conditions, ISD flux vacua must satisfy the tadpole bound
     \begin{equation} \label{eq:tadpole}
       0 < N_{\rm flux} < Q_{D3} 
\,,
    \end{equation}
    where 
        \begin{equation} \label{eq:nflux}
        N_{\rm flux}=\int_X H_3 \wedge F_3 = f^T \cdot \Sigma \cdot h \,,
    \end{equation}
    and $Q_{D3}$ encodes the D3-charge contribution induced from orientifold planes. 
    This bound on $N_{\rm flux}$ is essential for the finiteness of  the number of flux vacua, see e.g.~\cite{Ashok:2003gk,Acharya:2006zw,Denef:2004ze}.

There is one further important aspect for the phenomenological relevance of ISD vacua: the magnitude of the vacuum expectation value of the gauge-invariant  superpotential
\begin{equation}
    W_0 = \sqrt{2/\pi} \big< e^{K/2} W \big> \,.
\end{equation}
Targeted searches for type IIB vacua often puts a bound on $|W_0|$. In particular, this is true of leading proposals for the construction of de Sitter vacua: KKLT  \cite{Kachru:2003aw} requires that $|W_0|<<1$, whereas the LVS \cite{Balasubramanian:2005zx}  construction  requires $|W_0| \sim {\mathcal{O}}(1)$, in non-dimensional units.  We refer the interested reader to \cite{Cicoli:2013swa,Demirtas:2021nlu} for discussions of the relevance of $|W_0|$.    

\subsection{Field space geometry}    
The geometry of the Calabi-Yau complex structure moduli space is captured by the periods $\Pi_i(z)$, which are the holomorphic volumes of the CY 3-cycles.  With respect to a symplectic basis $(A_I,B^J)$ of $H_3(X,\mathbb{Z})$, intersecting as specified by the symplectic matrix
    \[
    \Sigma = \begin{pmatrix}
        {\mathbf{0}} & \mathbb{I} \\
        -\mathbb{I} & \mathbf{0} \\
    \end{pmatrix} \,,
    \]
    we define
    \begin{equation}
    \Pi =(X^I, {\cal F}_I)^T \,, \;  \mbox{ where } 
    X^I = \int_{A_I} \Omega \,, \; 
    {\cal F}_I = \int_{B^J} \Omega \,, \; 
\end{equation}
    and $\Omega$ is the holomorphic top form of the CY threefold. Here $X^I$ are homogeneous coordinates for the complex structure moduli space, and ${\cal F}_I$ are determined by the so-called prepotential $F(z)$.\footnote{Recall that CY moduli spaces are special K\"ahler manifolds; thus the real K\"ahler potential is determined by a holomorphic prepotential \cite{Strominger:1990pd,Candelas:1990pi}.} The periods satisfy a set of differential equations, the Picard-Fuchs equations, and are non-trivial functions of the complex structure moduli. Expressions of the periods, with limited radius of convergence, can be deduced once the Calabi-Yau geometry $X$ has been specified. These computations are non-trivial, and scale badly with $h^{(2,1)}$, the number of complex structure moduli of $X$. 
    
     The simplest expressions for the  periods are obtained in the region of large complex structure (LCS). We will focus on this region in this paper. In the LCS region, the prepotential decomposes into a classical part, which is polynomial in $z$, and non-perturbative corrections, which are exponentially suppressed. Both parts are computable from mirror symmetry: $F(z)$ can be deduced using topological information of the mirror dual $\widetilde{X}$ of the CY manifold $X$ \cite{Candelas:1990rm,Ceresole:1992su,Hosono:1993qy}:
    \begin{equation} \label{eq:prepot}
        F(z) = -\frac{1}{6} \tilde{\kappa}_{i j k} z^i z^j z^k + \frac{1}{2}a_{i j}z^i z^j+b_i z^i + \tilde{\xi} + F_{\rm inst}(z) \, ,
    \end{equation}
where $\tilde{\kappa}_{i j k}$ ($a_{i j},b_i$) are integer (rational) topological invariants of $\widetilde{X}$. 
The  constant $\tilde{\xi}$ is determined by the Euler characteristic of $\widetilde{X}$, $\tilde{\xi} = i \zeta(3) \chi(\widetilde{X})/(16 \pi^3)$. 
In the following, we will explore data for classical flux vacua, so we neglect the non-perturbative corrections to the prepotential.

We will need one more piece of information about the type IIB ${\cal N}=1$ four-dimensional supergravity, namely the symmetries of its field space. The 10D type IIB supergravity action is invariant under $SL(2;\mathbb{Z})$, which transform the axio-dilaton $\tau$ and flux vector $(f,h)$ as
\begin{equation}
    \tau \to \frac{a \tau + b}{c \tau + d} \, , \; 
    \begin{pmatrix}
        f \\ h
    \end{pmatrix} \to 
     \begin{pmatrix}
        a & b \\ c & d
    \end{pmatrix}
     \begin{pmatrix}
        f \\ h
    \end{pmatrix}
    \,, \; 
    \begin{pmatrix}
        a & b \\ c & d
    \end{pmatrix} \in SL(2,\mathbb{Z)} \,.
\end{equation}
$\nflux$ and $|W_0|$ are invariant under $SL(2;\mathbb{Z})$. 
Furthermore, the 4D theory is invariant under $Sp(2(h^{(2,1)}+1),2)$ transformations of the period and flux vectors, e.g. $\Pi \to M \Pi$, for $M \in Sp(2(h^{(2,1)}+1),2)$. These transformations correspond to monodromies of the 3-cycles, around special loci in the complex structure moduli space, which preserve the intersection matrix $\Sigma$ (i.e. $M^T \Sigma M = \Sigma$), as well as Kähler potential, superpotential, and $\nflux$.   We are particularly interested in the monodromy around the LCS point $z^i = 0$; these monodromies shift the real part of $z^i$ by integer values.

\subsection{Geometry used in experiments:  $\mathbb{CP}^4_{[11169]}[18]$}

For our experiments, we will explore a data set of ISD flux vacua generated for the study \cite{Chauhan:2025rdj}. The vacua arise from flux compactifications on
a CY threefold $X$, specified as a hypersurface in $\mathbb{CP}^4_{[11169]}$, given by the zero locus of the degree $18$ polynomial 
%\ml{Referee comment: swap $x_i$ to avoid confusion.}
%\begin{equation}
%    x_1^{18}+x_2^{18} + x_3^{18}+x_4^3+x_5^2-18\psi x_1 x_2 x_3 x_4 x_5 -3 \phi x_1^6 x_2^6 x_3^6 \,,
%\end{equation}
\begin{equation}
    w_1^{18}+w_2^{18} + w_3^{18}+w_4^3+w_5^2-18\psi w_1 w_2 w_3 w_4 w_5 -3 \phi w_1^6 w_2^6 w_3^6 \,,
\end{equation}
in the homogeneous coordinates $w_i$ of $\mathbb{CP}^4_{[11169]}$. This polynomial is invariant under a $\mathbb{Z}_6 \times \mathbb{Z}_{18}$ discrete symmetry, and identifies a complex structure moduli space of effective dimension $h^{(2,1)}=2$ for $X$. The two parameters   $\psi,\phi$ in the defining equation for $X$ are coordinates on this moduli space, whose relation to the affine coordinates $z^i$ will not be needed in the following (see \cite{Candelas_1994} for details). The $\mathbb{Z}_6 \times \mathbb{Z}_{18}$-invariant periods of $X$ were first computed in  \cite{Candelas_1994}, and the data needed for the prepotential \eqref{eq:prepot} is given by 
\begin{equation}
    \widetilde{\kappa}_{1,1,1} = 9 \,, \quad \widetilde{\kappa}_{1,1,2} = 9 \,, \quad \widetilde{\kappa}_{1,2,2} = 3\,, \quad a = \frac{1}{2} 
    \begin{pmatrix}
        9 & 3 \\
        3 & 0
    \end{pmatrix}\,, \quad b = \frac{1}{4} 
    \begin{pmatrix}
        17 \\
        6
    \end{pmatrix} \,, \quad \chi(\widetilde{X})= -540 \,.
\end{equation}
The relevant expressions for the non-perturbative corrections $F_{\rm inst}$ to the prepotential are known, and  were first computed in \cite{Candelas_1994}. As shown in \cite{Chauhan:2025rdj},  they are negligible for the vacua that we will study.

Even with the restriction to the classical prepotential, it is non-trivial to solve for ISD vacua in the LCS region of moduli space. Several properties conspire to make the problem hard to handle: the search space for integer flux vectors and continuous moduli is vast, making systematic searches costly; the LCS expansion for the prepotential \eqref{eq:prepot} is only valid in a restricted patch of the moduli space, and a given flux choice may not yield vacua in this patch; and numerical root-finding methods for polynomial equations typically require a good guess to converge to a solution at all. On the other hand, once a flux vector is given, and a point is selected in the moduli space, it is very easy to check if all conditions for vacua are met.

The computational complexity of the type IIB flux vacua motivates the present study. Using the results of  \cite{Chauhan:2025rdj}, our aim is thus to learn how to sample the 12-dimensional flux vectors ${\bf x}=(f, h)$  satisfying the ISD constraints \eqref{eq:Fterms} and the tadpole bound \eqref{eq:tadpole} with $Q_{D3}=276$. The vacua we train on have $N_{\rm flux}$ well below $Q_{D3}$, and lie in a region
\begin{equation} \label{eq:Utrain}
     \big\{ \Re({z^i}) \in (-0.5,0.5] \, , \,
    \Im({z^i}) \in [2,3] \, , \,
    c_0 \in (-0.5,0.5] \, , \,
    s \in \left[ \sqrt{3}/2,20\right]
    \big\} \,.
\end{equation}
which lies in intersection of the fundamental domain of the group $SL(2,\mathbb{Z})\times Sp(6,\mathbb{Z})$ and the LCS region, given by 
\begin{equation} \label{eq:U}
    U = \big\{ \Re({z^i}) \in (-0.5,0.5] \, , \,
    \Im({z^i}) > 1 \, , \,
    c_0 \in (-0.5,0.5] \, , \,
    s > \sqrt{3}/2 
    \big\} \,.
\end{equation}
Type IIB vacua that satisfy the F-term conditions \eqref{eq:Fterms}, the tadpole bound \eqref{eq:tadpole} and lie in region $U$ are called {\it valid} in the following.

\section{Generative models}
\label{sec:bgml}

In this paper, we use generative methods to learn the structure of type IIB flux vectors which yield ISD vacua satisfying certain criteria, as described in the previous section. In practise, we consider a set of  flux vectors $\mathbf{x} \in (x_1, \dots, x_{12})$, where $x_i \in \{\min, \dots, \max\}$, that make up the training data. Here $\min$ and $\max$ correspond to the minimum and maximum integer contained in the set respectively. These flux vectors correspond to valid type IIB vacua that lie in the region \eqref{eq:Utrain}, 
and hence can be seen as samples from the distribution of valid flux vacua 
\begin{equation}
    p_{\textrm{vacua}}(\mathbf{x})\,.
\end{equation}
We would like to draw new samples from this distribution. To this end, we utilize generative models as described in the introduction. In combination with a numerical solver, these trained models can be used to generate new type IIB ISD flux vacua.

In particular, we are using two specific generative models. The first is called Bayesian Flow Networks (BFNs) \cite{2023arXiv230807037G}. BFNs have a demonstrated ability to learn and sample from complex data distributions (like images), and {can also parse integer data. They} are thus expected to learn any non-trivial structure of the string landscape, when trained on a large set flux vectors that define valid string vacua. Additionally, BFNs are based on the method of Bayesian inference, making them a natural tool for modelling data distributions.

The second method uses a Transformer \cite{vaswani2023attentionneed} architecture. This also has the ability of learning structure of data, and generate new data samples (e.g. texts). In addition, a Transformer has the option of being prompted. This allows us to approximate conditional distributions of the type
\begin{equation}
    p_{\textrm{vacua}}(\mathbf{x}|y)\,,
\end{equation}
where $y$ can correspond to any additional condition that we wish the flux vacua to obey which leads to more control in the sampling process.  We will use Int2Int \cite{2025arXiv250217513C} to implement the Transformer architecture. In the following, we describe each model in more detail.

\subsection{Bayesian Flow Networks} \label{sec:bfn}

Bayesian Flow Networks are generative models that are closely related to Diffusion Models \cite{xue2024unifyingbayesianflownetworks}.\footnote{For a comprehensive overview of diffusion models, see \cite{yang2024diffusionmodelscomprehensivesurvey}.}  BFNs work for continuous and discrete data, the latter making them suitable for our problem at hand. In order to remain pedagogical, let us describe how the model works with a one-dimensional discrete example. The reader interested in a more detailed general discussion is referred to Appendix \ref{app:BFN_example}. 

Consider a set of one-dimensional data points whose distribution we are aiming to sample from, namely
\begin{equation}
    p_{\textrm{data}}(x) \,,
\end{equation}
where $x \in \{ 1, \dots, K \}$ for some positive integer $K$.\footnote{Type IIB fluxes will also take negative integer values. However, given a set of fluxes, one can shift all of them by $+|\min(x_i)|+1$ ensuring that the minimum value is $1$. During sampling, one then inverts this by subtracting the same value after the sampling process.} $x$ is distributed according to some unknown categorical distribution $\mathbf{p}_{\textrm{data}} = (p_1, \dots, p_K)$, where $p_k=p(k)$ corresponds to the probability that $x$ takes the value $k\in\{1, \dots, K\}$.

During the training process, four distributions are to be considered: the input, sender, receiver and output distribution. The input distribution strives to describe the data, conditioned on a categorical distribution $\mathbf{p}$, namely
\begin{equation}
    p_I(x|\mathbf{p})\,.
\end{equation}
During training, $\mathbf{p}$ will be initialized based on some information about the training data. This information consists of noised data samples drawn from the sender distribution, which is given by
\begin{equation}
    p_S\left(\mathbf{y}|x, \alpha(t)\right) \,,
\end{equation}
where $\alpha(t)=\beta(1)t$ is an accuracy parameter that measures how much noise is added, with $\alpha=0$ meaning that all information about $x$ is destroyed and $\alpha\to\infty$ meaning that all information is kept. $\beta(1)\in\mathbb{R}^+$ is some hyperparameter that can be tuned. 

Noised samples $\mathbf{y}$ are drawn for each data point $x$ at various times $t\in[0,1]$. These noised samples consist of the one-hot representation $\mathbf{e}_{x}$ of $x$ together with some added Gaussian noise, meaning one can infer the probability $\mathbf{p}$ that $\mathbf{y}$ was sampled conditioned on $x$ following some categorical distribution by $\mathbf{p}=\textrm{softmax}(\mathbf{y})$. This initial probability $\mathbf{p}$ is then fed into a neural network, together with time $t$, 
\begin{equation}
    \bm{\Psi}(\mathbf{p}, t) \,,
\end{equation}
which outputs an updated probability $\mathbf{p}_{\textrm{updated}}=\textrm{softmax}({\bm{\Psi}})$. This update is performed by minimizing a loss function for whose description, we need to introduce the receiver distribution
\begin{equation}
    p_R\left(\mathbf{y}|\mathbf{p}_{\textrm{updated}}, t, \alpha(t)\right) \,.
\end{equation}
The loss function is then given as the Kullback-Leibler divergence of the sender and receiver distribution
\begin{equation} \label{eq:kl_div}
    D_{KL}\left( p_S ||p_R \right) \,,
\end{equation}
which means that the receiver is trained to approximate the sender distribution, i.e. to learn how precisely the noise is added to the data, based on time $t$ and the updated probabilities received from the neural network.

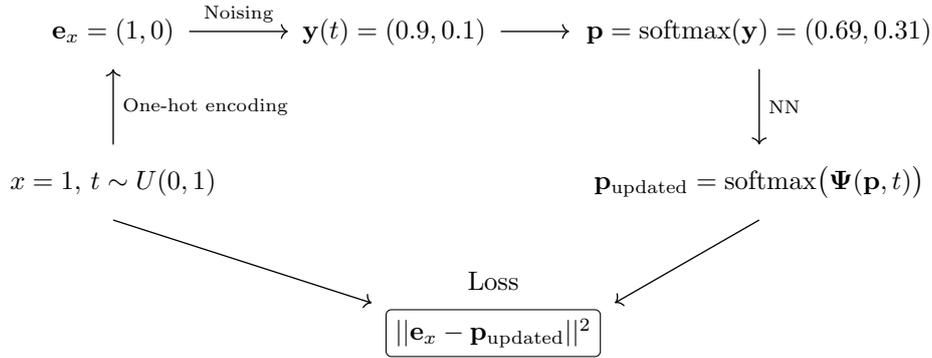
\begin{figure}
    \centering
    \begin{tikzpicture}
    \node[] at (0.2,4) {$\mathbf{y}(t)=(0.9, 0.1)$};
    \draw[->, black, line width=0.5pt] (-2.5, 4) -- node[above]{\scriptsize Noising} (-1.2, 4);
    \node[] at (-3.5, 4) {$\mathbf{e}_x=(1,0)$};
    \draw[->, black, line width=0.5pt] (-3.5, 2.5) -- node[right]{\scriptsize One-hot encoding} (-3.5, 3.5);
    \node[] at (-3.5, 2) {$x=1$, $t\sim U(0,1)$};
    \draw[<-, black, line width=0.5pt] (-0.1,0.4) -- (-3.5,1.5);
    \node[draw, rounded corners=2pt] at (1.5,0) {$|| \mathbf{e}_{x} - \mathbf{p}_{\textrm{updated}}||^2$};
    \node[] at (1.5, 0.7) {Loss};
    \draw[<-, black, line width=0.5pt] (3.1,0.4) -- (5,1.5);
    \draw[<-, black, line width=0.5pt] (5, 2.5) -- node[right]{\scriptsize NN} (5, 3.5);
    \node[] at (5, 2) {$\mathbf{p}_{\textrm{updated}}=\textrm{softmax}\big(\bm{\Psi}(\mathbf{p},t)\big)$};
    \draw[<-, black, line width=0.5pt] (2.5, 4) -- (1.6, 4);
    \node[] at (5, 4) {$\mathbf{p}=\textrm{softmax}(\mathbf{y})=(0.69, 0.31)$};
\end{tikzpicture}
    \caption{Schematic drawing of BFN training for the one-dimensional case with $K=2$. A training step is shown for one data point $x=1$.}
    \label{fig:bfn_training}
\end{figure}

Computing Eq. (\ref{eq:kl_div}) leads to the loss function
\begin{equation}
    \mathcal{L}(x) \sim || \mathbf{e}_{x} - \mathbf{p}_{\textrm{updated}}||^2 \,.
\end{equation}
In practise, this means that the network is trained to predict the data point $x$, based on the noised probability $\mathbf{p}$ and time $t$ it has received, i.e. it learns to remove noise from samples, where $t$ informs about the magnitude of noise added. A schematic drawing of this process is shown in Fig. \ref{fig:bfn_training}.

During sampling, this process it simply inverted. At the beginning of the sampling process, a completely noised sample together with time $t=0$ is fed into the network. $t$ is then gradually increased and at each new time step the sample is again fed into the network, removing more and more noise and finally outputting a fully detailed sample $x$.

\subsection{Transformers}
Transformer architectures underlie the large language models of popular chat bots like ChatGPT. There is an abundance of literature on these models, and the interested reader is referred to the seminal paper \cite{vaswani2023attentionneed} and the pedagogical review \cite{lin2021surveytransformers}.  The main benefits of Transformers exploited in this paper are the possibility of prompting the model, i.e. we can sample from them based on some condition, and their attention mechanism. Below, we give a rough overview of these features. More details regarding the implementation with Int2Int can be found in Appendix \ref{app:transformer}.

Transformers map sequences to sequences, and were originally developed for language translation tasks.
The architecture consists of an encoder and a decoder. The encoder receives a tokenized input\footnote{The tokenization is a fixed procedure, dependent on the training data. For example, in language generation the tokens can correspond to words in a dictionary. In mathematical settings, the tokens can correspond to integers in a certain range together with a token for the plus and minus sign respectively.} which is mapped into an embedding space. The resulting sequence of points in the embedding space, together with some positional encoding,\footnote{This positional encoding can be learned or be fixed to some function. It is a crucial feature that allows the encoder to understand the order of a sequence.} is fed into an attention layer followed by a feed forward layer. This is repeated for a fixed number of times, which produces the final encoder output. These recurrent steps allow the encoder to relate each element of the sequence to the other elements and learn any correlation that might exist, via the attention mechanism. 

The encoder output is then fed into the decoder to generate the desired output sequence. For this, an initial output token is generated, which in our case is the \textbf{start} token indicating the start of a sequence. This token is mapped into the same embedding space as for the encoder and then fed into an attention layer whose output is again fed into an attention layer together with the encoder output. As a last step, a feed forward layer is applied. Again, these steps are repeated a fixed number of times and as a final step, a linear layer is applied and the output is fed into a softmax function. This generates an output distribution which assigns a probability to each possible token. This procedure is needed to draw the next token, \textbf{token}$_0$, from the distribution
\begin{equation}
    p_{\bm{\theta}} \left( \textbf{token}_0 \big| \textbf{start}, \textbf{input} \right) \,,
\end{equation}
where $\bm{\theta}$ corresponds to the network parameters, and \textbf{input} corresponds to the input fed into the encoder. Once the next token is drawn, it is again fed into the decoder, together with the previous token and the decoder output. This step can be repeated several times, where each token is drawn from a probability distribution conditioned on the encoder input and the previously generated sequence. This can be summarized in the following distribution
\begin{equation}
    p_{\bm{\theta}} \left( \textbf{token}_i \big| \textbf{token}_{i-1}, \textbf{token}_{i-2}, \dots, \textbf{start}, \textbf{input} \right) \,,
\end{equation}
where $\textbf{token}_i$ corresponds to the token drawn at the $i$th repetition of the decoder 
step. This is repeated until either the sequence reaches a maximum length or a token is drawn that indicates the end of the sequence. The resulting sequence of tokens is then mapped to the desired output, e.g. a sentence or a sequence of integers. 

During training, the model is optimized to match corresponding sequences in the training data, e.g. mapping sentences from one language into another. The trained model can then be used to draw new sequences based on an input sequence, i.e. one can draw from the conditional distribution
\begin{equation}
    p_{\bm{\theta}}( \textbf{output}  |  \textbf{input})\,.
\end{equation}
In our experiments, the output sequences are integer sequences with fixed length, namely the flux vectors $\mathbf{x}$ corresponding to valid ISD vacua. The input sequences are one-dimensional, i.e. a single integer, and correspond to certain properties that we want those vacua to have. Say this property is a specific $\nflux$ that those vacua should compute, then training a Transformer yields the conditional distribution 
\begin{equation}
    p_{\bm{\theta}} ( \mathbf{x} | \nflux) 
\end{equation}
from which samples can be drawn by prompting the Transformer with the respective value of $\nflux$.

\section{Experiments}
\label{sec:expIIB}

In this section, several experiments are performed to test the performance in sampling flux vectors of both BFNs and Transformers. 
To train the models, the data\footnote{The dataset can be found on \url{https://github.com/ml4physics/JAXvacua}.} generated in Example A in \cite{Chauhan:2025rdj} is used. In the following, this dataset is referred to as \textit{Dataset A}. This dataset contains $\sim 5~000~000$ fluxes in the region (\ref{eq:Utrain}) with minimum and maximum value of $-86$ and $86$ respectively. 

After training, the models are used to sample fluxes,  for which the F-term conditions are solved numerically, using SciPy's \verb|root| method.\footnote{The numerical tolerance is selected to be $10^{-5}.$ Selecting this tolerance resulted in rather fast validity checks, leveraging JAX's just-in-time compilation. Selecting a smaller tolerance or choosing an initial guess far away from the solution leads to poorer performance which constitutes a numerical limitation of this method.} The resulting moduli and axio-dilaton, together with the fluxes, are then mapped by appropriate SL$(2, \mathbb{Z})$ and monodromy transformations into the region $\Re({z^i}) \in (-0.5,0.5]$, $c_0 \in (-0.5,0.5]$ and $|\tau|>1$. Finally, the results are checked for validity.\footnote{As mentioned previously, a sample is referred to as valid, if it solves Eq. (\ref{eq:Fterms}) and leads to moduli $z^1$, $z^2$ and axio-dilaton $\tau=c_0+is$ in the region $U$, as defined in Eq. (\ref{eq:U}). }
 For the BFN experiments, the condition $\nflux>0$ is also checked, which is not necessary for the Transformer experiments.
Since the Transformers are prompted with certain values, their capability of producing vacua that correctly compute said values is also checked. For both methods, the capability of interpolation/extrapolation, i.e. the amount of samples that are not contained in the training data, is also studied.\footnote{As mentioned in the introduction, we consider as new solutions any solution not contained in the training data, i.e. some subset of Dataset A. The authors of \cite{Chauhan:2025rdj} claim to be exhaustive in the region (\ref{eq:Utrain}), thus we cannot expect our solutions to give rise to genuinely new vacua. However, we sample many solutions outside of said region and can safely assume that those are not contained in Dataset A (they may still be contained in the other datasets listed in \cite{Chauhan:2025rdj}).} The trained models, together with some example evaluation code, can be found on \url{https://github.com/mowalden/sampling_flux_vacua}.

Throughout this work, we use the following terminology. \textit{Validity} refers to the ratio of valid solutions among all (numerically unique) samples that were generated by the trained models. \textit{Originality} refers to the fraction of valid samples that are not contained in the training data. In the Transformer experiments, \textit{Accuracy} refers to the fraction of samples that compute the correct value (i.e. the value that was used in the prompt) among valid ($|W_0|$) or numerically unique ($\nflux$) samples.

\subsection{Sampling vacua using Bayesian Flow Networks} \label{sec:BFN_experiment}

\begin{figure}
    \centering
    \includegraphics[width=\linewidth]{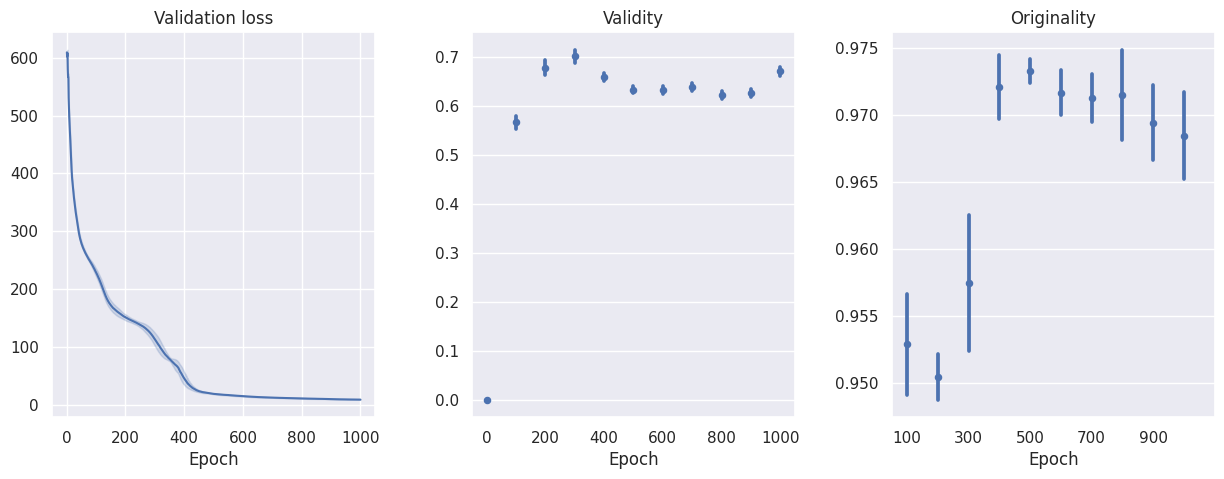}
    \caption{Training evaluation of the BFNs. The validation loss, validity and originality are plotted over various epochs. The values correspond to the mean of five models that have been trained on different subsets of Dataset A and the errors correspond to the standard deviation. The validity corresponds to the percentage of valid fluxes among numerically unique samples. Originality corresponds to the percentage of valid fluxes not contained in the training data.}
    \label{fig:bfn_training_CP11169}
\end{figure}

We apply BFNs to the aforementioned setting in order to evaluate the capability of the architecture to sample new valid fluxes. More precisely, we are aiming to sample from the distribution of valid flux vacua by approximating it with a BFN, namely 
\begin{equation}
     p_{\bm{\theta}}(\mathbf{x}) \approx p_{\textrm{vacua}}(\mathbf{x})\,,
\end{equation}
where $\bm{\theta}$ corresponds to the network parameters and $\mathbf{x} = (x_1,\dots,x_{12})$, $x_i\in \{-86, \dots, 86\}$, are the fluxes that we wish to sample.

We are interested in the interpolation and extrapolation abilities of the method, meaning how well certain distributions are reproduced and how well additional solutions, not contained in the training data, are found. Furthermore, we are interested in whether any structure can be learned by the models at all, since the space of valid flux vacua is rather small compared to the space of $12$-dimensional integer vectors.\footnote{For a $12$-dimensional vector with integer values ranging from $-86$ to $86$, the number of different combinations is $\mathcal{O}(10^{185})$. Thus even the smallest validity in sampling, confirms that a model has learned some non-random structure in the space of flux vectors.}

In the following BFN experiments, a randomly sampled subset of Dataset A, consisting of $1~000~000$ data points, containing all possible $\nflux \in \{4, \dots, 34 \}$, with flux numbers ranging from $-86$ to $86$, is used. During training the data is split into $50\%$ training and $50\%$ validation data. Five models\footnote{For more details regarding hyperparameters, see Appendix \ref{app:BFN_hyperparameters}.} are trained on different randomly sampled datasets for $1~000$ epochs each. The models are then evaluated by considering their mean validity with standard deviation as an error measure. 

In order to evaluate the trained models, $10~000$ samples with flux numbers in the range of $-86$ to $86$ are generated at every 100th epoch and the subset of numerically unique fluxes is kept.\footnote{The size of this subset was always $\sim10~000$ which is why this measure is dropped from the following evaluation.} Randomly sampled integers in the range of $-86$ to $86$ serve as a benchmark. The validity of those random samples always amounted to $0\%$ and is hence dropped from the following evaluations. 

In Fig. \ref{fig:bfn_training_CP11169}, the validation loss, validity and originality during training is shown. At epoch 300, the models sample with a validity of $\sim 0.7$, which means that $\sim 70\%$ of sampled fluxes give valid solutions,\footnote{For the remaining $30\%$, either SciPy's \verb|root| does not find any solution ($\sim20\%)$ or the solution is outside of the region of LCS and/or leads to invalid axio-dilaton values ($\sim10\%$). The condition $\nflux>0$ is almost always met. In the cases in which no solution is found, we expect the solution to lie outside the region of LCS, far away from the initial guesses that are used in the \verb|root| method, which illustrates a numerical limitation rather than the non-existence of a solution.} and then converges to an validity of $\sim 0.65$. Furthermore, the originality of samples, i.e. how many of the generated fluxes are not contained in the training data, is checked. This amounted to $\sim 95\%$, attesting the models excellent interpolation capabilities.  

\begin{figure}
    \centering
    \includegraphics[width=\linewidth]{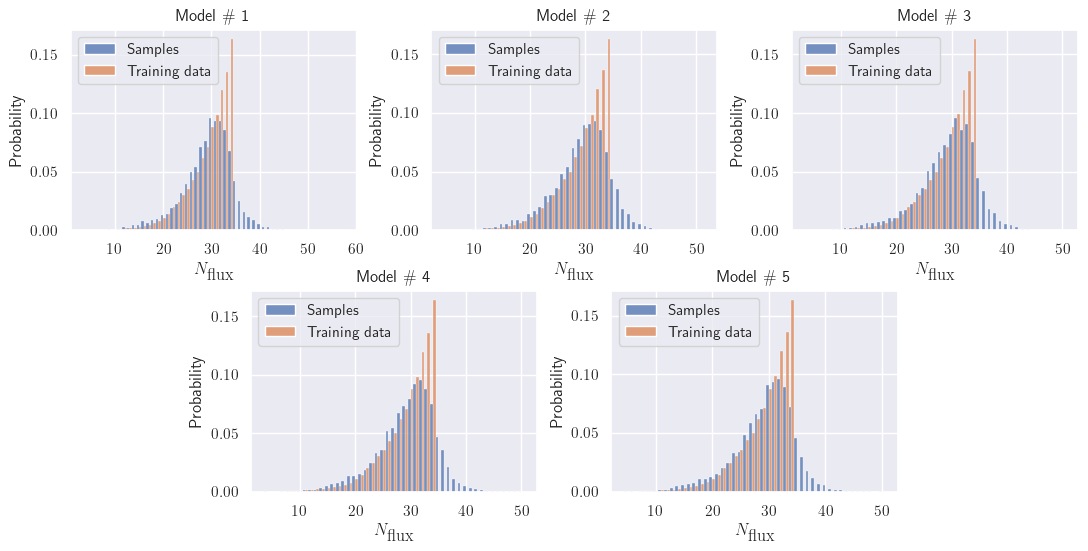}
    \caption{Distribution of $N_\textrm{flux}$ computed from samples drawn from all five models at epoch 300 compared to the respective training data. The plotted samples correspond to the valid samples and thus make up $\sim70\%$ of the $10~000$ samples.}
    \label{fig:N_distribution_BFN_CP11169}
\end{figure}

\begin{figure}[]
    \centering
    \includegraphics[width=\linewidth]{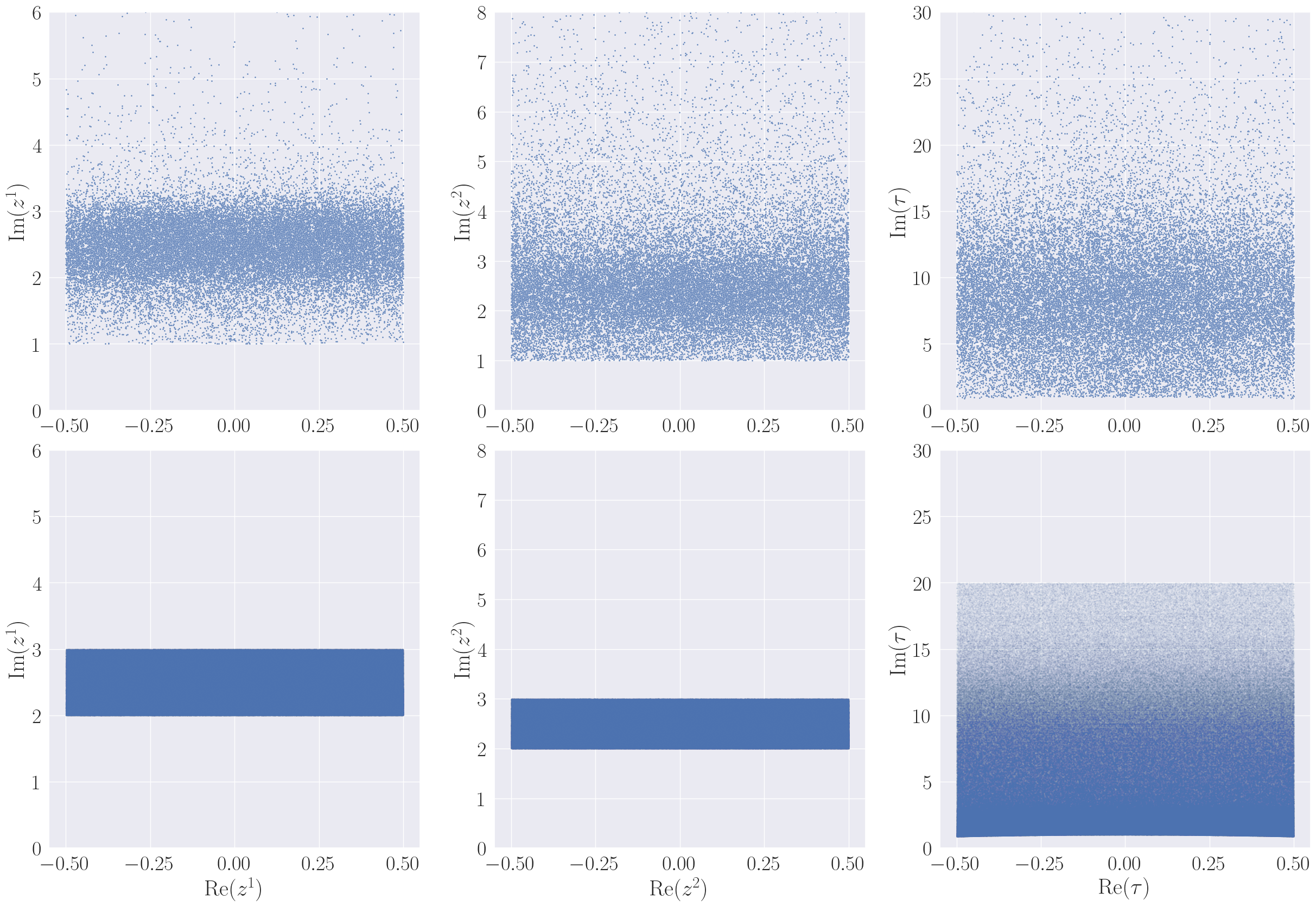}
    \caption{Distribution of $z^1, z^2$ and $\tau$ computed from valid samples drawn from five models at epoch 300 (first row) compared to the training data (second row). Points in each plot correspond to samples/training data from all five models. To ensure readability, the sample plots have been cut off resulting to some samples not being shown.}
    \label{fig:moduli_distribution_BFN_CP11169}
\end{figure}

Besides the validity of the samples, we are also interested in how well the trained models reproduce 
certain distributions. We start with the distribution of $\nflux$ for which $10~000$ fluxes are sampled for each model, which are then filtered by validity. The $\nflux$ distribution of the remaining $70\%$ of samples is plotted in Fig. \ref{fig:N_distribution_BFN_CP11169} and compared to the respective training data. It can be observed that the BFNs reproduce the data distribution up to $\nflux=30$ rather well but perform worse for $\nflux > 30$. Remarkably, roughly 10\% of the fluxes sampled by the models have $\nflux>34$, values for which no examples were contained in the training data, attesting them extrapolation capabilities.\footnote{The distribution of $W_0$ and the moduli of those fluxes is identical to the distributions in Fig. \ref{fig:moduli_distribution_BFN_CP11169} and \ref{fig:distribution_W0}.}

\begin{figure}[]
    \centering
    \includegraphics[width=0.9\linewidth]{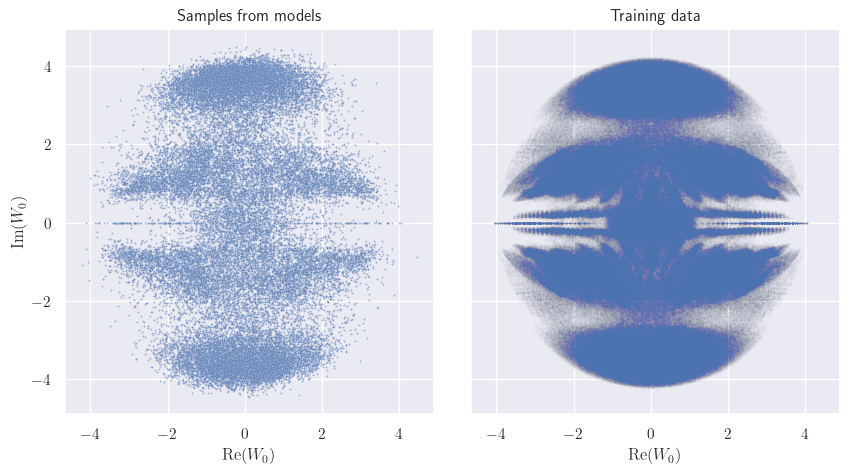}
    \caption{Distribution of $W_0$ computed from samples drawn from five models at epoch 300 compared to the training data (points in each plot correspond to valid samples/training data from all five models).}
    \label{fig:w0_distribution_BFN_CP11169}
\end{figure}

Furthermore, we are interested in the distribution of the moduli $z^i$, the axio-dilaton $\tau$ and $W_0$. In particular, we would like to see similar distributions as observed in \cite{Chauhan:2025rdj}. In Fig. \ref{fig:w0_distribution_BFN_CP11169}, the distribution of $W_0$ obtained from samples is plotted and compared with the training data. An adequate match of the distributions can be observed. In particular, high density regions in the training data are captured well by the trained models. Of particular interest is the region $|W_0|$ close to zero, which is populated both by the training data and the samples. If one aims to sample mostly samples for which $|W_0|$ is close to zero, this can be achieved by an adequate preparation of the training data. However, in this experiment we are interested in reproducing the overall distribution of $W_0$ that is seen in the training data.

The plots for $z^i$ and $\tau$ are given in Fig. \ref{fig:moduli_distribution_BFN_CP11169}. Again, it can be observed that the distributions of the training data are captured well by the trained models. The models reproduce the region of the training data as high density regions and are also capable of extrapolating outside of that domain. For $\tau$, the training data shows a high concentration of values towards the lower edge of the fundamental domain which does not seem to be captured by the trained models.

\subsection{Conditional sampling of vacua using Transformers}
As a second experiment, we apply a Transformer architecture to the aforementioned setting. The clear advantage over the BFN is that Transformers can be conditioned. A disadvantage is longer training time and that the implementation is more involved.

In this section, two different conditions are explored: $\nflux$ and $|W_0|$. We are training Transformers to learn the distribution of flux vacua that either compute a specific $\nflux$ or lead to solutions that lie in a certain range of $|W_0|$. In the following, these two cases are explored in detail.

\subsubsection{Conditioning on $\nflux$}\label{sec:nflux}

We aim to sample from a conditional distribution of valid flux vacua by approximating it with a Transformer, namely
\begin{equation}
    p_{\bm{\theta}} (\mathbf{x}|\nflux) \approx p_{\textrm{vacua}}(\mathbf{x}|\nflux) \,,
\end{equation} 
where $\bm{\theta}$ corresponds to the network parameters and $\mathbf{x} = (x_1,\dots,x_{12})$, $x_i \in \{-100, \dots, 100\}$,  are the fluxes that we wish to sample.\footnote{The range of flux integers is slightly expanded to comply with the Int2Int package.} This can be straightforwardly implemented using the Transformer architecture as described in the following.

A set of $500~000$ fluxes of Dataset A is randomly sampled and prepared by matching each $\textbf{x}$ with the corresponding $\nflux \in \{4, 5, \dots, 34\}$. The remaining fluxes of the dataset are used as a validation set. A single model\footnote{For details regarding hyperparameters, see Appendix \ref{app:transformer_hyperparameters}.} was trained for $2~000$ epochs and its sampling performance is evaluated. 

\begin{figure}[!t]
    \centering
    \includegraphics[width=\linewidth]{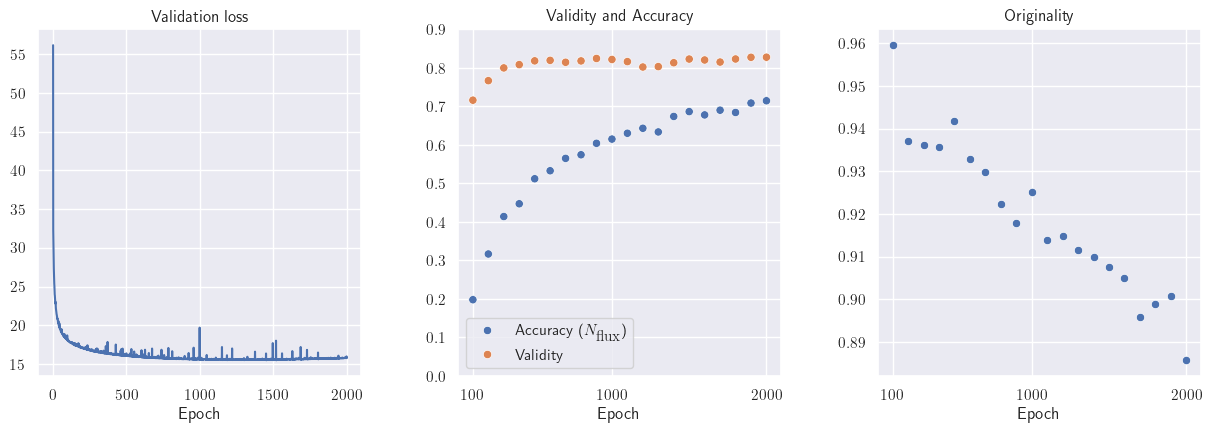}
    \caption{Training evaluation of a Transformer conditioned on $\nflux$. The validation loss, validity, accuracy (for $\nflux=24$) and originality are shown. Accuracy measures the percentage of samples that compute the correct $N_{\textrm{flux}}$ among a numerically unique subset of the $10~000$ (i.e. $\#\textrm{valid}/\#\textrm{numerically unique}$). The percentage of samples that are valid solutions among the samples that compute the correct $N_{\textrm{flux}}$ are also reported. Finally, the originality of samples is plotted.}
    \label{fig:transformer_trainng_CP11169.png}
\end{figure}

In Fig. \ref{fig:transformer_trainng_CP11169.png}, the validation loss and the accuracy, as well as the originality of samples, are plotted at every 100th epoch. For this, $10~000$ samples are generated and the subset\footnote{In the better performing regions, the size of numerically unique samples was close to $10~000$. However, in the regions that performed worse, these subsets were significantly smaller (see Table \ref{tab:N-accuracy_transformer_CP11169}).} of numerically unique solutions is evaluated. Accuracy reflects the ability of the model to sample fluxes with correct $\nflux$. This is done by prompting the model with $\nflux=24$ and then checking whether the sampled flux vector computes the correct $\nflux$. After filtering out the samples with $\nflux\neq 24$, the validity is checked and then the originality, i.e. the percentage of valid and accurate fluxes not contained in the training data, of the samples is plotted. Fig. \ref{fig:transformer_trainng_CP11169.png} shows a high originality of $>88\%$, attesting the architecture very good interpolation capabilities. Validity for accurate $\nflux$ remains stable at $\sim 80\%$ during training. The accuracy starts low but reaches a high value of $\sim 70\%$ towards the end of training.

\begin{table}[!t]
    \centering
    \resizebox{0.6\textwidth}{!}{
    \begin{tabular}{|c|c|c|c|c|}
        \hline
        \bm{$N_\textrm{flux}$} & \makecell{\textbf{Numerically} \\ \textbf{unique samples}} & \textbf{Accuracy} & \textbf{Validity} & \textbf{Originality}  \\
        \hline
         4 & 467 & 7.49 \% & 0.00 \% & -- \\
         \hline 
         5 & 154 & 16.23 \% & 68.00 \% & 88.23 \% \\
         \hline 
         6 & 537 & 45.25 \% & 65.84 \% & 91.87 \% \\
         \hline 
         7 & 1241 & 34.41 \% & 39.11 \% & 90.42 \% \\
         \hline 
         24 & 9790 & 73.78 \% & 82.09 \% & 89.73 \% \\
         \hline 
         25 & 9783 & 73.67 \% & 83.02 \% & 88.52 \% \\
         \hline 
         26 & 9839 & 72.01 \% & 84.20 \% & 89.30 \% \\
         \hline 
         27 & 9861 & 72.05 \% & 84.67 \% & 89.31 \% \\
         \hline
         
    \end{tabular}
    }
    \caption{Evaluation of samples obtained by conditioning a Transformer on $\nflux$. Samples consist of $10~000$ fluxes for each prompted $N_{\textrm{flux}} \in\{4, 5, 6, 7, 24, 25, 26, 27\}$. Accuracy measures how many numerically unique samples compute the correct $N_{\textrm{flux}}$ and validity measures the percentage of accurate samples that are also valid and unique up to SL$(2,\mathbb{Z})$ and monodromy transformations. The last column measures the originality of accurate and valid samples.}
    \label{tab:N-accuracy_transformer_CP11169}
\end{table}

\begin{figure}
    \centering
    \begin{minipage}{0.45\textwidth}
    \centering
    \includegraphics[width=\linewidth]{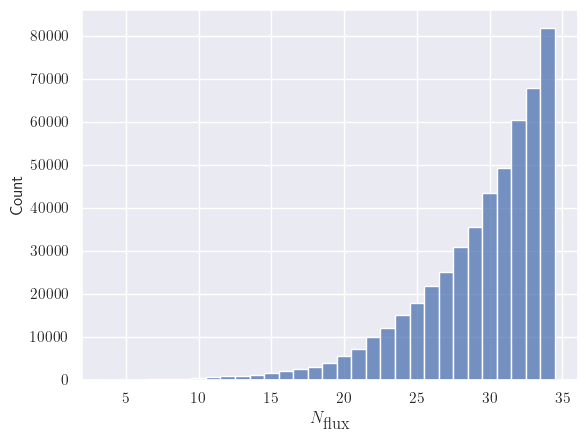}
        \caption{Distribution of fluxes with respect to $N_{\textrm{flux}}$ in the training data used for the Transformer architecture.}
\label{fig:data_dist_N_transformer_CP11169}
    \end{minipage}
    \hspace{1em}
    \begin{minipage}{0.45\textwidth}
           \centering
    \vspace{2.5em}
    \includegraphics[width=\linewidth]{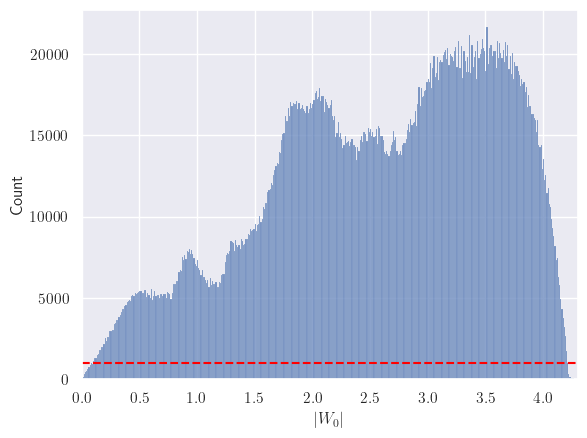}
    \caption{Distribution of Dataset A with respect to values of $|W_0|$. In order to ensure a more balanced training set, the number of samples per range has been limited to $1~000$ (red line). Close to zero, less than $1~000$ samples were available, leading to neglect of that region during training.}
    \label{fig:distribution_W0}
    \end{minipage}

\end{figure}

In Table \ref{tab:N-accuracy_transformer_CP11169}, the sampling capabilities of the model after training are listed for the regions of $\nflux$ in which it performs worst and best, i.e. $\nflux \in \{4, \dots, 7\}$ and $\nflux \in \{ 24, \dots, 27\}$ respectively. A table with all $\nflux \in \{ 4, \dots, 34\}$ can be found in Appendix \ref{app:nflux}. It can be observed that the model performs much better for larger $\nflux$ where it demonstrates very good sampling capabilities both with respect to the validity of vacua and the accuracy of computing the correct $\nflux$. In the region of low $\nflux$, adequate to poor performance is observed. This can be explained by the distribution of the training data (see Fig. \ref{fig:data_dist_N_transformer_CP11169}). More precisely, the skew in the training data towards larger $\nflux$ has the effect that the model focuses on those examples during training as opposed to examples with lower $\nflux$. Hence the Transformer sees larger $\nflux$ more frequently during training and thus performs better in that region. This effect could potentially be decreased by a more balanced dataset.

Prompting the network with  values outside of the region $\nflux \in \{4, \dots, 34 \}$ led to poor performance with an accuracy of nearly zero percent, since those values were not contained in the training data. This shows that, despite good interpolation abilities, the model is not capable of extrapolating very well with respect to the $\nflux$ distribution (as opposed to the BFN).

\begin{figure}
    \centering
    \includegraphics[width=0.9\linewidth]{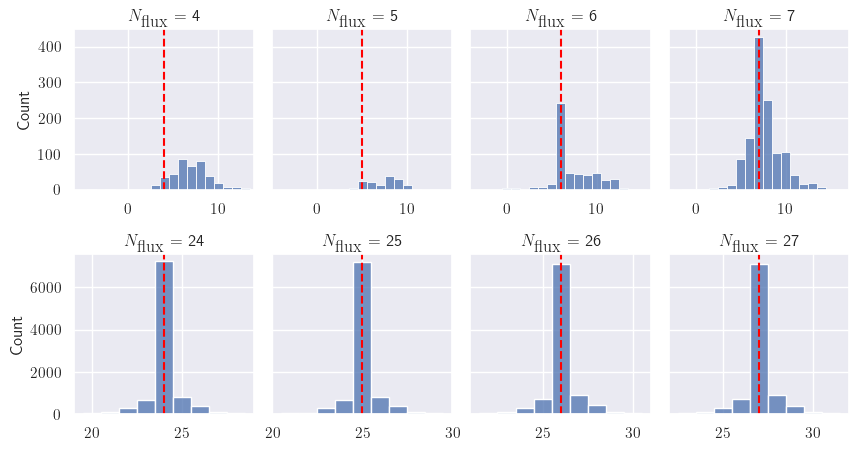}
    \caption{Sampling accuracy of a Transformer conditioned on $\nflux$. Samples consist of $10~000$ fluxes for each prompted $N_{\textrm{flux}} \in\{4, 5, 6, 7, 24, 25, 26, 27\}$. Numerically identical samples are removed and the remaining fluxes are used to compute the actual value of $N_{\textrm{flux}}$. The red line corresponds to the prompted value and the bins reflect the distribution of samples around it. For better readability, the range of the x-axis has been fixed, leading to some outliers not being shown in the plots.}
    \label{fig:N-accuracy_transformer_CP11169}
\end{figure}

In Fig. \ref{fig:N-accuracy_transformer_CP11169}, the distribution of $\nflux$ is plotted for the best and worst performing regions of $\nflux$, i.e. $\nflux \in \{4, \dots, 7\}$ and $\nflux \in \{ 24, \dots, 27\}$. A figure with all $\nflux \in \{ 4, \dots, 34\}$ can be found in Appendix \ref{app:nflux}. In the region of larger $\nflux$, very good performance is observed while adequate to poor performance can be observed in the lower region. However, despite the low accuracy in regions of lower $\nflux$, the ground truth value is still the most sampled for all $\nflux>5$.

\subsubsection{Conditioning on $|W_0|$}

Another interesting property to consider is the expectation value of the gauge-invariant GVW-superpotential $|W_0|$. To this end, we train a Transformer to approximate and sample from a conditional distribution of flux vacua, namely
\begin{equation}
    p_{\bm{\theta}} (\mathbf{x}\big||W_0|) \approx p_{\textrm{vacua}} (\mathbf{x}\big||W_0|)\,,
\end{equation}
where again $\bm{\theta}$ corresponds to the network parameters and $\mathbf{x} = (x_1,\dots,x_{12})$, $x_i \in \{-100, \dots, 100\}$, are the fluxes that we wish to sample.
Dataset A is sorted by values of $|W_0|$ and $1~000$ randomly drawn samples are selected for each interval in the set \linebreak $|W_0| \in\{ [0, 0.01], (0.01,0.02], (0.02, 0.03], \dots, (3.99, 4] \}$. For values up to $0.1$, less than $1~000$ samples are available and hence all examples are selected. This gives rise to a slightly skewed dataset (see Fig. \ref{fig:distribution_W0}), leading to less exposure of the Transformer to values $|W_0|\leq 0.1$ during training. The remaining fluxes in Dataset A are used as validation data. 

\begin{figure}
    \centering
    \includegraphics[width=\linewidth]{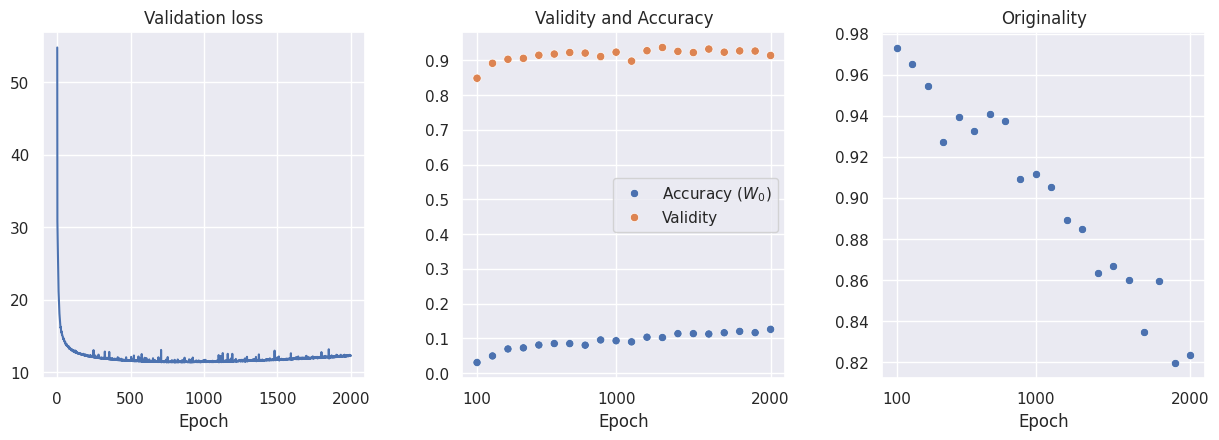}
    \caption{Training evaluation of a Transformer conditioned on $|W_0|$. The validation loss, validity, accuracy and originality are shown for $|W_0|=4.0$. Validity measures the percentage of samples among a numerically unique subset of $10~000$ samples that give rise to valid solutions (i.e. $\#\textrm{valid}/\#\textrm{numerically unique}$). Accuracy measures the percentage among valid samples that also compute the correct $|W_0|$. Finally, the originality of samples is plotted.}
    \label{fig:transformer_training_W0}
\end{figure}

A single model with the same hyperparameters as in Sec. \ref{sec:nflux} is trained for $2~000$ epochs. The validation loss, validity, accuracy and originality are shown in Fig. \ref{fig:transformer_training_W0}. High validity of $\sim 90\%$ during the entirety of training and an accuracy of $\sim 10\%$ from epoch $1~000$ onwards can be observed. However, slight overfitting is taking place after epoch $1~000$ which can be measured through an increasing validation loss and decreasing originality, hence the model is evaluated at epoch $1~000$ in the following.

\begin{figure}
    \centering
    \includegraphics[width=0.7\linewidth]{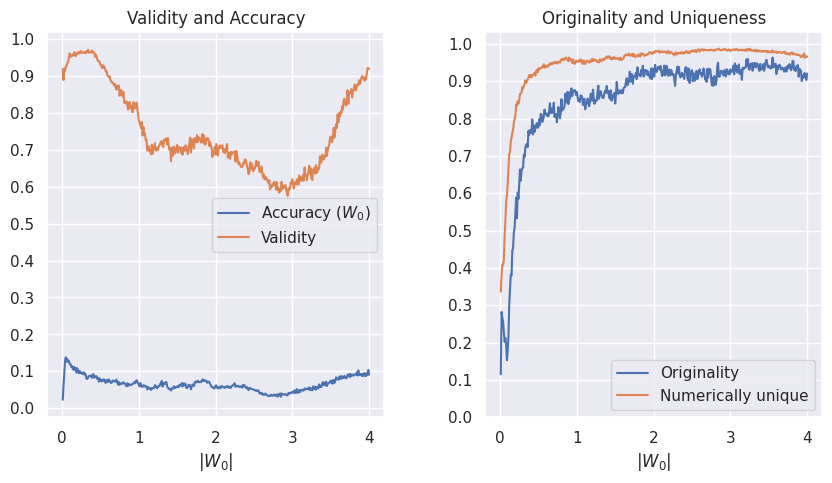}
    \caption{Evaluation of samples obtained by conditioning a Transformer on $|W_0|$. Samples consist of $10~000$ fluxes for each prompted $|W_0|\in\{ [0,0.01], (0.01,0.02], \dots, (3.99, 4] \}$. Only numerically unique samples are kept (shown in right figure) and checked for validity (shown in left figure). The resulting moduli and axio-dilaton are used to compute $|W_0|$ to check whether the solution lies in the desired $|W_0|$ range (this is labelled as accuracy). Furthermore, the interpolation capability of the model is reported (this is labelled as originality).}
    \label{fig:transformer_evaluation_W0}
\end{figure}

Fig. \ref{fig:transformer_evaluation_W0} shows the performance of the Transformer over the entire range of $|W_0|$ that was used during training. In the region of $|W_0|$ close to zero, the model shows high accuracy but low validity, meaning that most solutions do not fall into the range of e.g. $|W_0| \in [0,0.01]$. Furthermore, only a small subset of the $10~000$ samples are numerically unique. However, it is still possible to sample new vacua in this regime. We managed to sample 9, 68 and 102 new\footnote{By new, we mean that the flux vectors were not contained in the training data. The lowest value found is $|W_0|\approx 0.0026$ (the lowest value in the training data is $|W_0|\approx 0.0018$).} vacua for the regions $|W_0|\in[0,0.01]$, $|W_0|\in(0.01,0.02]$ and $|W_0|\in(0.02, 0.03]$. In better performing regions, e.g. $|W_0|$ close to $4$, significantly better performance is observed which results in a higher number of samples, namely up to $\sim 900$ new vacua. This discrepancy can be explained by the skew of the training data and the fact that only a small set of examples with low $|W_0|$ were available upon training. Fig. \ref{fig:transformer_evaluation_W0} shows several dips in accuracy and validity for larger $|W_0|$ and a drastic performance increase in the region of $|W_0|\sim 4$. This can be explained by looking at the distribution of the training data in Figs. \ref{fig:w0_distribution_BFN_CP11169} and \ref{fig:distribution_W0}. Even though, the data has been filtered to make the training set more balanced, the distribution indicates that it is significantly easier to find solutions in high $|W_0|$ regions, while a dip occurs at $\sim2.5$. This behaviour seems to be reflected in the performance of the trained Transformer. One would assume that this leads to poor accuracy performance in regions of low $|W_0|$ as well. However, looking at originality and uniqueness in this region, we can see that the Transformer samples only a small subset of numerically unique and original solution which indicates that the high accuracy performance is due to the fact that it samples mostly fluxes that are seen in the training data.

\begin{figure}
    \centering
    \includegraphics[width=\linewidth]{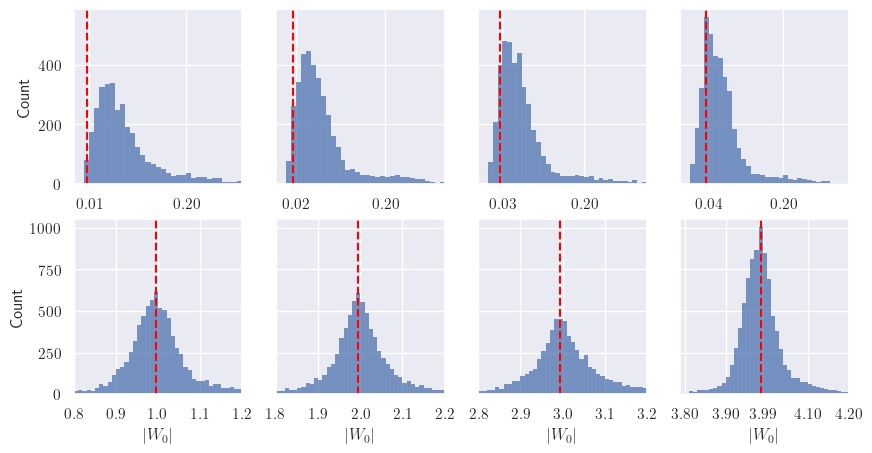}
    \caption{Sampling accuracy of a Transformer conditioned on $|W_0|$. Samples consist of $10~000$ fluxes for each prompted $|W_0|\in\{ [0, 0.01], (0.01,0.02], (0.02, 0.03], (0.03, 0.04], (0.99, 1], (1.99,2], (2.99,3], (3.98,3.99] \}$. The numerically unique subset is checked for validity. For the valid solutions, $|W_0|$ is computed and the results are plotted in intervals of width $0.01$. The red line corresponds to the prompted value and the bins reflect the distribution around it. For better readability, the range of the x-axis has been fixed, leading to some outliers not being shown in the plots.}
    \label{fig:transformer_W0_accuracy}
\end{figure}

In Fig. \ref{fig:transformer_W0_accuracy}, the distribution of $|W_0|$ around a selection of prompted values is plotted.\footnote{Prompting a specific value $x$ means that the interval $|W_0|\in(x-0.01, x]$ is considered.} It can be observed that for $|W_0|$ close to zero, the Transformer fails to sample the correct value in most cases but the solutions do cluster in a region close to the prompted value. At $|W_0|\in(0.03,0.04]$, the model starts to become more accurate. For the remaining higher values, the prompted value is the most often sampled, however a spread of non-negligible width can is observed.

\section{Discussion and outlook}
\label{sec:d&o}
In this paper, we explored sampling techniques for type IIB flux vacua using generative models. To this end, two models have been explored: Bayesian Flow Networks and Transformers. Both models have shown very good performance in the task of generating flux vectors that give rise to type IIB flux vacua. 

The trained models were used to sample flux vectors for which the F-term conditions \eqref{eq:Fterms} were solved numerically. If a solution was found, it was checked whether the resulting moduli $z^i$ and the axio-dilaton $\tau$ lay in the region defined in Eq. \eqref{eq:U}. If the tadpole bound \eqref{eq:tadpole} was also satisfied, a vacuum was found. Generative models are approximate by nature, which is a needed property to enable finding new solutions and not just simple reproduction of the training data. Thus, the aforementioned validity checks are a crucial part of our workflow. This is different to "exact" ML methods, like Reinforcement Learning and Genetic Algorithms, where the output solutions already satisfy all specified constraints. Different methods should still be pursued in order to understand how they compare.

In our experiments, Bayesian Flow Networks proved apt at sampling type IIB flux vacua. We achieved an accuracy of nearly $70\%$, as opposed to $0\%$ accuracy when fluxes where drawn at random, where almost all generated flux vectors gave rise to original vacua, i.e. vacua that were not contained in the training data. The model also managed to reproduce the distribution of vacua with respect to $\nflux$, $|W_0|$, moduli $z^i$ and axio-dilaton $\tau$. We also observed minor extrapolation capabilities where the method produced around 10\% sampled vacua outside of the training region. A drawback of this model, is that it is unconditional, meaning that fluxes cannot be drawn based on some prompt.

For conditional sampling, we used a Transformer, which showed good performance in finding new vacua that satisfy additional constraints. These constraints were chosen to either be a specific value for $\nflux$ or an interval for $|W_0|$. Both examples produced good results but the conditioning on $\nflux$ worked significantly better, which was to be expected since this condition is independent of the field space coordinates. Furthermore, $|W_0|$ is a continuous variable and the discretization was somewhat arbitrary, i.e. if one were to choose wider intervals, the performance would obviously increase. The trained Transformers demonstrated very good interpolation capabilities, i.e. $\sim80-90\%$ of sampled flux vectors gave rise to original vacua. As expected, no significant extrapolation capabilities were observed. In both cases, we observed that the performance is very much influenced by the distribution of the training data. In general, one needs to be able to generate large numbers of configurations to act as training data which can prove difficult in settings where such generations are not, or only to limited extent, possible. To circumvent this problem, one could implement a similar fine-tuning process as in \cite{Yip:2025hon}, e.g. to attain more examples of small $|W_0|$. We hope to come back to this issue in a future publication.

With the aforementioned results, these methods prove to be powerful tools in the exploration of type IIB string vacua. Furthermore, those methods could be combined with packages like JAXvacua \cite{Dubey:2023dvu} and CYtools \cite{Demirtas:2022hqf} to create even more versatile programs. However, since this is a method paper, we are not pushing the boundary of string model building. To do this, one needs a more thorough study and preparation of the setup for the problem at hand. I.e. if we want to find vacua with $|W_0|$ close to zero, we would change our data set and employ the previously mentioned fine-tuning process. Furthermore, to achieve concrete small values for $|W_0|$, the numerical precision of all used methods has to be studied carefully.

While we have optimized our models, it is possible that e.g. sample accuracy could be improved by further hyperparameter tuning.  Another interesting direction to explore is conditional sampling in the context of BFNs, similar to conditional sampling with Diffusion Models \cite{ho2022classifierfreediffusionguidance}.

It would also be interesting look at other geometries in the type IIB setting, i.e. with large $h^{2,1}$. Scaling behaviour of both performance and training time could be investigated. Furthermore, other settings in string theory would be of interest, e.g. heterotic line bundle standard models scans \cite{Anderson:2011ns,Anderson:2012yf,Larfors:2020ugo, Abel:2023zwg}. We hope to come back to this in a future publication.

\section*{Acknowledgements}
This research was in part supported by  Vetenskapsrådet, grant no. 2020-03230. 
The computations were partially enabled by resources provided by the National Academic Infrastructure for Supercomputing in Sweden (NAISS), partially funded by the Swedish Research Council through grant agreement no. 2022-06725.
MW thanks Carlos Rodriguez for initial collaboration in the project.
We thank Anna Dawid, Jim Halverson, Ziwei Luo, Ayca Özcelikkale, Fabian Ruehle, Andreas Schachner and Arno Solin for valuable discussions. Special thanks to Andreas Schachner for explanations of the datasets of ref. \cite{Chauhan:2025rdj}, and to Jim Halverson for suggesting to use ref. \cite{2025arXiv250217513C}.

%%%%%%
\clearpage
\appendix

\section{Bayesian Flow Networks}\label{app:BFN}

In this paper, we use Bayesian Flow Networks \cite{2023arXiv230807037G} (BFNs) to perform the experiments in Sec. \ref{sec:BFN_experiment}. The code used both for training and sampling is taken from \url{https://github.com/nnaisense/bayesian-flow-networks/}.

In Sec. \ref{app:BFN_example}, we illustrate the training and sampling procedure for BFNs. Furthermore, we list the hyperparameters used for our experiments in Sec. \ref{app:BFN_hyperparameters}.

\subsection{Example}\label{app:BFN_example}

Let us illustrate the procedure by using the concrete example of  vectors in $\mathbb{Z}^n$ (which may correspond to the flux vectors in ISD vacua of type IIB string theory). Clearly, this amounts to a discrete setting, namely our data consists of points $\mathbf{x} = (x_1, \dots, x_D) \in \{ 1, \dots, K \}^{D}$, i.e. we consider $D$ integers which can take values between $1$ and $K$. Then we have for each $x_i$ a vector of class probabilities $\mathbf{p}^{(i)}=(p_1^{(i)}, \dots , p_K^{(i)})$, where $p_k^{(i)}\in [0,1]$ is the probability that $x_i$ has value $k$. Thus we can combine the categorical probabilities for all $x_i$ in a $KD$-dimensional vector $\mathbf{P} = (\mathbf{p}^{(1)}, \dots, \mathbf{p}^{(K)})$. 

We would like to be able to capture some correlation between the different distributions. For example, say we want to sample series of consecutive integers $\mathbf{x} = (x_1, \dots, x_D)$, then the categorical distribution of $x_i$ clearly influences the distribution of $x_{i+1}$. Namely, if the most likely value for $x_i$ is some integer $k$, the most likely value for $x_{i+1}$ should be $k+1$ and so on. In the BFN, this dependence is modelled by a neural network.

In more detail, this works as follows. Let $\mathbf{e}_{\mathbf{x}}=(e_{x_1}, \dots, e_{x_D})\in \mathbb{R}^{KD}$ be the one-hot encoding of $\mathbf{x}$, such that $(e_{x_i})_k=\delta_{x_i k}$. The underlying categorical distribution of $\mathbf{x}$ is never fed into the network directly. Instead a noisy version of $\mathbf{x}$ is sampled, which is used by the network to infer information about $\mathbf{x}$, i.e. it learns to remove noise from a given sample. The noising together with the inference step can be combined into the Bayesian Flow Distribution\footnote{Hence the name Bayesian Flow Network. For derivation of the distribution, see \cite{2023arXiv230807037G}.}
\begin{equation}
    p_F= \underset{\mathcal{N}(\mathbf{y}|\beta(t)(K\mathbf{e}_\mathbf{x}-\mathbf{1}), \beta(t) K \mathbb{I})}{\mathbb{E}} \delta (\mathbf{P} - \textrm{softmax}(\mathbf{y})) \,,
\end{equation}
where $\beta(t)=t^2 \beta(1)$, $\beta(1) \in \mathbb{R}^+$, is the accuracy schedule that controls the magnitude of noise added. Note that we have introduced the time variable $t$ which tells us how much noise was added. 

During training, we consider $t\in[0,1]$ as a continuous variable and at each epoch sample $t \sim U(0,1)$. We sample from $p_F$, by first drawing
\begin{equation}
    y\sim \mathcal{N}(\mathbf{y}|\beta(t)(K\mathbf{e}_\mathbf{x}-\mathbf{1}), \beta(t) K \mathbb{I})  \; ,
\end{equation}
and then setting $\mathbf{P}=\textrm{softmax}(\mathbf{y})$.
 These updated parameters are then, together with $t$, fed into the network 
 \begin{equation}
     \Psi(\mathbf{P}, t) = (\Psi^{(1)}(\mathbf{P}, t), \dots, \Psi^{(D)}(\mathbf{P}, t)) \; ,
 \end{equation}
 where $\Psi^{(i)}(\mathbf{P},t) = (\Psi^{(i)}_1(\mathbf{P},t), \dots, \Psi^{(i)}_K(\mathbf{P},t))$. The network output is then turned into class probabilities which we define as the output distribution
\begin{equation}
    p_O^{(i)}(k|\mathbf{P},t) = \left( \textrm{softmax}(\Psi^{(i)}(\mathbf{P},t)) \right)_k \,,
\end{equation}
which denotes the probability that variable $x_i$ will take value $k$. From this, we construct
\begin{equation}
    \hat{\mathbf{e}}^{(i)} (\mathbf{P},t) = \sum_{k=1}^K p_O^{(i)}(k|\mathbf{P},t) \mathbf{e}_k \,,
\end{equation}
where $(\mathbf{e}_k)_j = \delta_{kj}$. After defining
\begin{equation}
    \hat{\mathbf{e}}(\mathbf{P},t) = (\hat{\mathbf{e}}^{(1)} (\mathbf{P},t), \dots, \hat{\mathbf{e}}^{(D)} (\mathbf{P},t))\,,
\end{equation}
we define the loss
\begin{equation}\label{eq:loss}
    L(\mathbf{x}) = K \beta(1) \, t \, || \mathbf{e}_{\mathbf{x}}- \hat{\mathbf{e}}(\mathbf{P},t)||^2 \,.
\end{equation}
This loss is derived from the Kullback-Leibler divergence of the sender distribution to the receiver distribution which we have defined heuristically for the one-dimensional case in Sec. \ref{sec:bfn}. For a detailed derivation of those distributions in the general case, see \cite{2023arXiv230807037G}. 

Intuitively, one can think of Eq. (\ref{eq:loss}) as training the network to de-noise the updated class probabilities it has received.

Once the network is trained, we can use it to sample new data points. We start with a uniform prior $\mathbf{P}=(\frac{1}{K}, \dots, \frac{1}{K}) \in [0,1]^{KD}$. At the first step, we have $t=0$ and sample $k_i \sim \textrm{softmax} (\Psi^{(i)}(\mathbf{P},t))$, such that $\mathbf{k}=(k_1, \dots, k_D)$. These output probabilities are then used to generate a noisy sample 
\begin{equation}
    y\sim \mathcal{N}(\alpha(K\mathbf{e}_{\mathbf{k}}-\mathbf{1}), \alpha K \mathbb{I}) \; ,
\end{equation}
where $\alpha(t) = \beta(1) 2 t$ is the accuracy parameter. This noisy sample is then used to update the class probabilities, such that 
\begin{equation}
    \mathbf{P}_{\textrm{updated}} = \frac{e^{\mathbf{y}}\mathbf{P}}{\sum_k (e^{\mathbf{y}}\mathbf{P})_k} \,.
\end{equation}
We then set $\mathbf{P} = \mathbf{P}_{\textrm{updated}}$ and repeat the process for a given number of steps until the final sample is drawn from $k_i \sim \textrm{softmax} (\Psi^{(i)}(\mathbf{P},t=1))$.

\subsection{Hyperparameters}\label{app:BFN_hyperparameters}
The network consists of an input layer of width $2076$, eight hidden layers of width $2048$ and an output layer of width $2076$, all with ReLu activation function. A batchsize of $10~000$ was chosen. All hidden layers and the final layer are preceded by dropout layers with dropout probability of $20\%$ for the hidden layers and $50\%$ for the final layer. As an optimizer, we use AdamW with a learning rate of $0.0001$, $\lambda=0.01$, $\beta_1=0.9$ and $\beta_2 = 0.98$. $\beta(1)=3$, which is needed for the accuracy schedule $\beta(t)=t^2\beta(1)$. The following hyperparameters were optimized with respect to sampling accuracy: 
 \begin{itemize}
     \item Optimizer Adam with and without weight decay,
     \item Learning rate $\in \{ 10^{-3}, 10^{-4}, 10^{-5} \}$,
     \item With and without dropout,
     \item $\beta(1) \in \{0.3, 3.0, 10.0\}$,
     \item Number of hidden layers $\in \{ 1, 2, 3, 5, 8\}$,
     \item Width of hidden layers $\in \{ 512, 1024, 2048\}$.
 \end{itemize}
 All runs were performed on an NVIDIA A40 GPU\footnote{We have also observed adequate performance on smaller networks that can be trained on less sophisticated hardware within reasonable time.} with 48 GB RAM and took approximately one and a half hours each for $1~000$ epochs.

\section{Transformer architecture}
\label{app:transformer}

To analyse mathematical problems with Int2Int \cite{2025arXiv250217513C}, the first step is to assemble the training data in a format that the model can parse. In the case at hand, the training data are flux vectors, i.e.~sequences of integers, which lead to type IIB flux vacua. In our experiments, Dataset A is transformed according to the requirements of Int2Int. That means that fluxes are saved in a text file where each row corresponds to the input sequence (i.e. $\nflux$ or $|W_0|$), a \verb|<TAB>|, \verb|V12| to indicate the beginning of a sequence with 12 entries, and the flux integers with their respective signs. For example, the flux vector
\begin{equation}
    \mathbf{x} = (-2, 13, -2, 0, -1, 5, 2, -2, -2, 0, 1, -2) \,,
\end{equation}
with $\nflux=25$, is saved as
\begin{verbatim}
          25<TAB>V12 - 2 + 13 - 2 + 0 - 1 + 5 + 2 - 2 - 2 + 0 + 1 - 2
\end{verbatim}
in the text file. Int2Int then loads and tokenizes those integer sequences, and trains a Transformer architecture on the given data. The package has several components, as described in Sec. 1.3 in \cite{2025arXiv250217513C}. We will use most of these components with no, or very minor, modifications. In particular, the data-loader, model, trainer and optimizer, will be used in their original form, taken from \url{https://github.com/f-charton/Int2Int}, commit from Oct 7 2024.

In order to evaluate the trained networks, we slightly modify Int2Int's evaluator. Instead of prompting with input from an evaluation file, we prompt it with specific values of $\nflux$ or $|W_0|$ and sample output with a sampling temperature of $1.0$. The generated fluxes are then saved and evaluated using our custom code.

In Sec. \ref{app:transformer_hyperparameters}, we listed the hyperparameters used for training. In Sec. \ref{app:nflux}, additional plots of the performance evaluation are shown.

\subsection{Hyperparamters}\label{app:transformer_hyperparameters}
Hyperparameters were adopted from similar mathematical problems \cite{2025arXiv250217513C, halverson2025learningtopologicalinvariance}. A thorough hyperparameter study using a grid search was not feasible due to long runtimes and limited resources. The following hyperparamters were chosen:
\begin{itemize}
    \item Adam with learning rate $10^{-4}$,
    \item Encoding layers: $4$,
    \item Decoding layers: $4$,
    \item Embedding dimension: $256$,
    \item Encoding heads: $8$,
    \item Decoding heads: $8$,
    \item Hidden layers encoder: $1$,
    \item Hidden layers decoder: $1$,
    \item Batch size of $10~000$.
\end{itemize}

The runs were performed on an NVIDIA A40 GPU with 48 GB RAM and took roughly $24$ hours each for $2~000$ epochs. Samples were generated with sampling temperature $1.0$.

\subsection{Conditioning on $\nflux$ -- Additional plots}\label{app:nflux}

In this section, we present additional plots for the experiments in Sec. \ref{sec:nflux}.

In Tab. \ref{tab:N-accuracy_transformer_CP11169_appendix}, the sampling performance of the trained Transformer is shown over the entire training range of $\nflux$. 

\begin{figure}
    \centering
    \includegraphics[width=0.9\linewidth]{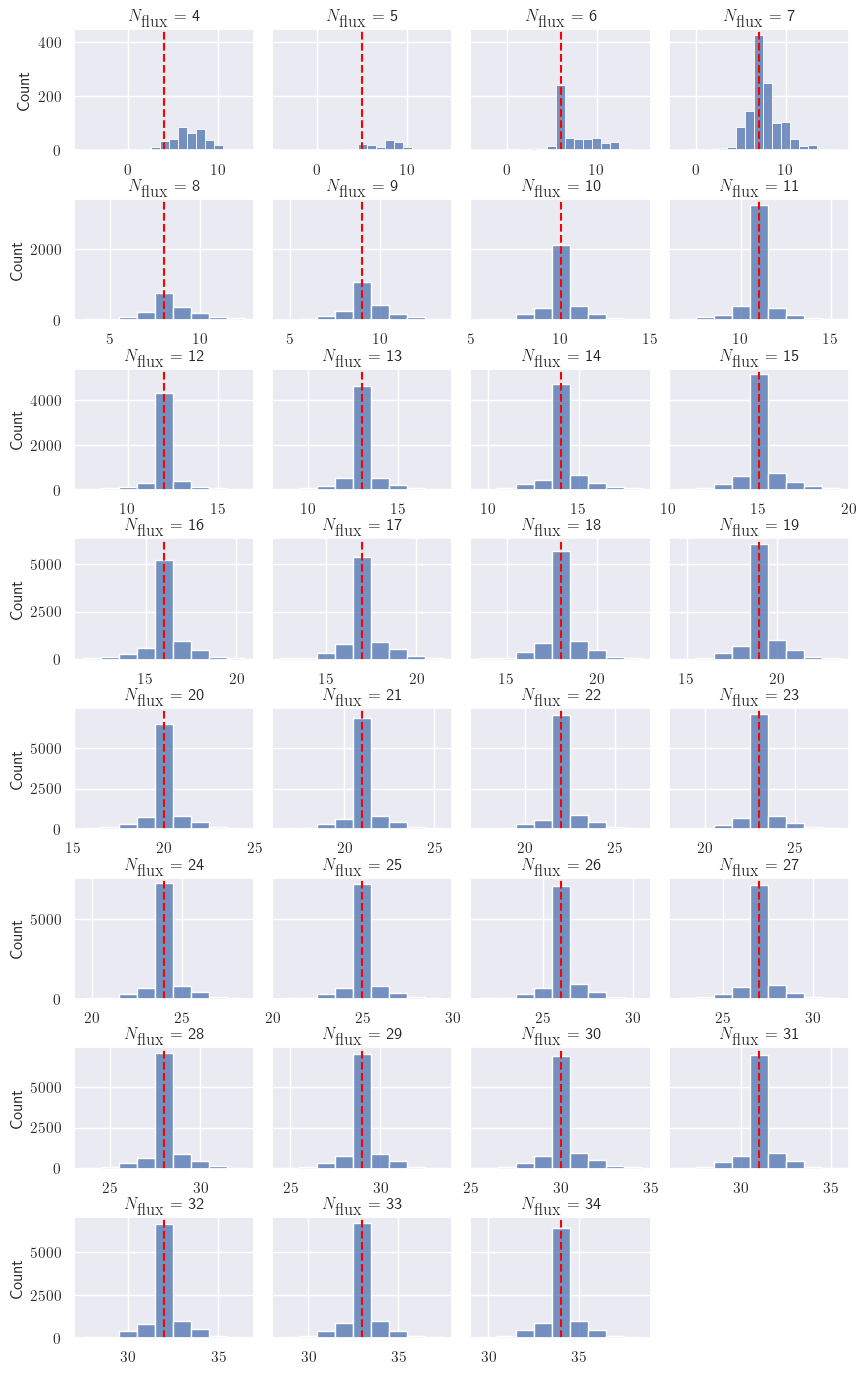}
    \caption{Sampling accuracy of a Transformer conditioned on $\nflux$. Samples consist of $10~000$ fluxes for each prompted $N_{\textrm{flux}} \in\{4, \dots, 34\}$. Numerically identical samples are removed and the remaining fluxes are used to compute the actual value of $N_{\textrm{flux}}$. The red line corresponds to the prompted value and the bins reflect the distribution of samples around it. For better readability, the range of the x-axis has been fixed, leading to some outliers not being shown in the plots.}
    \label{fig:N-accuracy_transformer_CP11169_appendix}
\end{figure}

In Fig. \ref{fig:N-accuracy_transformer_CP11169_appendix}, the distribution of $\nflux$ is plotted for all $\nflux$ contained in the training data. 

\begin{table}
    \centering
    \resizebox{0.6\textwidth}{!}{
    \begin{tabular}{|c|c|c|c|c|}
        \hline
        \bm{$N_\textrm{flux}$} & \makecell{\textbf{Numerically} \\ \textbf{unique samples}} & \textbf{Accuracy} & \textbf{Valid} & \textbf{Originality}  \\
        \hline
         4 & 467 & 7.49 \% & 0.00 \% & -- \\
         \hline 
         5 & 154 & 16.23 \% & 68.00 \% & 88.23 \% \\
         \hline 
         6 & 537 & 45.25 \% & 65.84 \% & 91.87 \% \\
         \hline 
         7 & 1241 & 34.41 \% & 39.11 \% & 90.42 \% \\
         \hline 
         8 & 1994 & 38.06 \% & 33.20 \% & 87.70 \% \\
         \hline 
         9 & 2303 & 47.07 \% & 49.17 \% & 87.80 \% \\
         \hline 
         10 & 3572 & 59.35 \% & 61.27 \% & 86.14 \% \\
         \hline 
         11 & 4580 & 71.24 \% & 76.34 \% & 84.79 \% \\
         \hline 
         12 & 5637 & 76.64 \% & 80.28 \% & 83.56 \% \\
         \hline 
         13 & 6478 & 71.47 \% & 78.27 \% & 85.71 \% \\
         \hline 
         14 & 6928 & 68.49 \% & 77.72 \% & 84.76 \% \\
         \hline 
         15 & 7656 & 67.42 \% & 78.90 \% & 84.34 \% \\
         \hline 
         16 & 8227 & 63.55 \% & 72.25 \% & 91.87 \% \\
         \hline 
         17 & 8584 & 62.34 \% & 70.17 \% & 84.39 \% \\
         \hline 
         18 & 8893 & 63.94 \% & 70.47 \% & 86.10 \% \\
         \hline 
         19 & 9138 & 66.06 \% & 73.45 \% & 87.10 \% \\
         \hline 
         20 & 9349 & 69.42 \% & 75.53 \% & 88.23 \% \\
         \hline 
         21 & 9493 & 72.10 \% & 76.80 \% & 88.32 \% \\
         \hline 
         22 & 9614 & 73.53 \% & 79.56 \% & 88.20 \% \\
         \hline 
         23 & 9685 & 73.53 \% & 80.28 \% & 88.21 \% \\
         \hline 
         24 & 9790 & 73.78 \% & 82.09 \% & 89.73 \% \\
         \hline 
         25 & 9783 & 73.67 \% & 83.02 \% & 88.52 \% \\
         \hline 
         26 & 9839 & 72.01 \% & 84.20 \% & 89.30 \% \\
         \hline 
         27 & 9861 & 72.05 \% & 84.67 \% & 89.31 \% \\
         \hline 
         28 & 9891 & 71.39 \% & 84.66 \% & 89.80 \% \\
         \hline 
         29 & 9888 & 71.19 \% & 84.77 \% & 89.76 \% \\
         \hline 
         30 & 9937 & 69.22 \% & 85.61 \% & 90.25 \% \\
         \hline 
         31 & 9922 & 69.88 \% & 86.77 \% & 89.44 \% \\
         \hline 
         32 & 9941 & 67.03 \% & 85.19 \% & 90.26 \% \\
         \hline 
         33 & 9951 & 67.52 \% & 84.91 \% & 90.04 \% \\
         \hline 
         34 & 9957 & 64.38 \% & 85.34 \% & 89.32 \% \\
         \hline
         
    \end{tabular}
    }
    \caption{Evaluation of samples obtained by conditioning a Transformer on $\nflux$. Samples consist of $10~000$ fluxes for each prompted $N_{\textrm{flux}} \in\{4, \dots, 34\}$. Accuracy measures how many numerically unique samples compute the correct $N_{\textrm{flux}}$ and validity measure the percentage of accurate samples that are also valid and unique up to SL$(2,\mathbb{Z})$ and monodromy transformations. The last column measures the originality of accurate and valid samples.}
    \label{tab:N-accuracy_transformer_CP11169_appendix}
\end{table}

%%%%%%%%%%%%%%%%%%%%
\newpage
\bibliographystyle{bibstyle}

\bibliography{bibliography}

\end{document}